\documentclass[footinbib,twocolumn,amsmath,amstex,amssymb,mathfonts,superscriptaddress,prl]{revtex4-1}
\usepackage[draft]{graphicx}
\usepackage[usenames, dvipsnames]{color}
\usepackage{bm}
\usepackage{verbatim}
\usepackage{amsmath}
\usepackage{amssymb}
\usepackage{amsthm}
\usepackage{amsfonts}
\usepackage{soul}

\def\red{\textcolor{red}}

\newcommand{\ad}[1]{\text{ad}_{S_{#1}(t)}}

\usepackage{bbm}

\begin{document}

\title{Floquet Mechanism for Non-Abelian Fractional Quantum Hall States}

\author{Ching Hua Lee}
\affiliation{Institute of High Performance Computing, A*STAR, Singapore, 138632.}
\email{phylch@nus.edu.sg}
\affiliation{Department of Physics, National University of Singapore, Singapore, 117542.}
\author{Wen Wei Ho} 
\affiliation{Department of Physics, Harvard University, Cambridge, Massachusetts 02138, USA}
\author{Bo Yang} 
\affiliation{Complex Systems Group, Institute of High Performance Computing, A*STAR, Singapore, 138632.}
\author{Jiangbin Gong}
\affiliation{Department of Physics, National University of Singapore, Singapore, 117542.}
\author{Zlatko Papi\'c}
\affiliation{School of Physics and Astronomy, University of Leeds, Leeds LS2 9JT, UK}
\email{z.papic@leeds.ac.uk}


\date{\today}
\begin{abstract}
Three-body correlations, which arise between spin-polarized electrons in the first excited Landau level, are believed to play a key role in the emergence of enigmatic non-Abelian fractional quantum Hall (FQH) effects. Inspired by recent advances in Floquet engineering, we investigate periodic driving of anisotropic two-body interactions as a route for controllably creating and tuning effective three-body interactions in the FQH regime. 
We develop an analytic formalism to describe this Floquet-FQH protocol, which is distinct from previous approaches that instead focus on bandstructure engineering via modulation of single-particle hopping terms. 
By systematically analyzing the resulting interactions using generalized pseudopotentials, we show  that our Floquet-FQH approach leads to repulsive as well as attractive three-body interactions that are highly tunable and support a variety of non-Abelian multicomponent FQH states.
Finally, we propose an implementation of the protocol in optically dressed ultracold polar molecules with modulated Rabi frequencies.

\end{abstract}

\maketitle 

Topological phases exhibit enticing prospects for 
fault-tolerant quantum computation~\cite{freedman2003topological,nayak2008non}
owing to their exotic quasiparticle excitations~\cite{Moore1991362, KitaevHoneycomb, Read-PhysRevB.61.10267}. 
These phases are believed to arise from an interplay between the Coulomb interaction, Landau level quantization and complete spin polarization in 2D electronic systems~\cite{Prange}, as suggested by the observation of even-denominator FQH plateaus in semiconductors~\cite{Willett87} and recently in bilayer graphene~\cite{Morpurgo2013, Zibrov}. The unexpected even-denominator plateaus are explained by   adiabatic continuity~\cite{Morf1998,HaldaneRezayi2000,Storni} between the underlying gapped many-electron state and the ground state of a system with special 3-body electronic interactions~\cite{Greiter92,ReadRezayiZeroModes}. Such 3-body interactions condense the electrons into a strongly-correlated quantum state where they fractionalize into non-Abelian Ising anyons~\cite{Moore1991362}. More generally, multi-body interactions are anticipated to give rise to other types of non-Abelian anyons~\cite{PhysRevB.82.245301,ardonne2009domain,barkeshli2010classification}.

Conventionally, effective 3-body interactions arise due to
Coulomb interactions and virtual excitations between Landau levels (LLs)~\cite{BisharaNayak,PetersonNayak,sodemann2013landau,simon2013landau,ghazaryan2017light}, 
a process suppressed by the LL splitting in a magnetic field, given by   the cyclotron energy $\hbar\omega_c = \frac{\hbar e B}{mc}$.
At the same time, the    incompressibility gap, which determines the stability of a FQH state, scales as $e^2/\epsilon \ell_B$, where $\ell_B=\sqrt{\hbar/eB}$ is the magnetic length. Thus, the effect of 3-body interactions can typically only be enhanced at the expense of reducing the energy gap, which weakens the FQH state.

Inspired by recent progress in ``Floquet engineering"~\cite{CayssolReview,bukov2015universal,eckardt2017colloquium,wang2018floquet,PhysRevLett.121.036401}, we propose an alternative method to realize effective 3-body interactions and hence stabilize
various non-Abelian FQH states.
Our approach consists of periodically modulating (2-body) interactions, specifically the repulsion between spatially separated electrons.
Key to our idea is the non-commutativity of the Girvin-MacDonald-Platzman (GMP) algebra~\cite{GMP85,GMP86} describing the   electron density operators projected to a LL, which   is the defining property of both continuum FQH states~\cite{GMP85} and   their lattice analogs, the fractional Chern insulators (FCIs)~\cite{ParameswaranRoySondhi}. We show that, owing to this algebra, the effective, {\it static} Hamiltonian that arises when a generic anisotropic FQH system is driven at high frequencies contains a rich set of many-body interactions which scale with the inverse driving frequency, rather than the LL gap. In particular,  desired 3-body multicomponent (spin) interactions can be engineered by time modulation of realistic 2-body interactions. More generally, we systematically analyze the interactions resulting from our ``Floquet-FQH" protocol using the framework of generalized pseudopotentials~\cite{yang2017generalized}, and show that the drive can also generate \emph{attractive} 3-body interactions. 

Finally, we discuss a realistic implementation of the Floquet-FQH protocol in ultracold molecules optically dressed with modulated Rabi frequencies, whose static version was previously established to host FCI states~\cite{yao2013realizing}. 
We  note that our approach is conceptually different from previous Floquet proposals~\cite{grushin2014floquet,anisimovas2015role,klinovaja2016topological,yap2018photoinduced} and experiments~\cite{jotzu2014experimental} which focused on topological band engineering via modulation of (single-body) kinetic terms; it is also distinct from   works~\cite{gong2009many,rapp2012ultracold,greschner2014density, meinert2016floquet}, which modulated \emph{on-site} 2-body interactions to probe tunnelling phenomena and Mott/superfluid phases.

{\it Two key inspirations:--}(i) A defining feature of FQH systems~\cite{GMP85} is the GMP algebra
\begin{equation}
[\bar \rho_\mathbf{q}^\sigma,\bar \rho_{\mathbf{q'}}^{\sigma'}]=2i\delta_{\sigma,\sigma'}\sin\frac{\hat{\mathbf{z}} \cdot (\mathbf{q}\times \mathbf{q'} ) \ell_B^2}{2}\bar \rho^\sigma_{\mathbf{q}+\mathbf{q'}},
\label{commutation}
\end{equation}
obeyed by the density operators $\bar\rho_\mathbf{q}^{\sigma} \equiv \mathcal{P}\rho_\mathbf{q}^\sigma \mathcal{P}=\sum_j e^{i\mathbf{q}\cdot \mathbf{R}_j^{\sigma}}$, projected to a given LL via $\mathcal{P}$.  Here, $R_j^{\sigma, a} \equiv r_j^{\sigma,a}+\ell_B^2\epsilon^{ab}\Pi^{\sigma}_{j,b}$ denote the guiding-center coordinates  of $j$-th particle with spin $\sigma$~\cite{Prange}, $\epsilon^{ab}$ is the antisymmetric tensor ($a=x,y$). 
Note that $R_j^{\sigma, a}$ differ from the position coordinates $r_j^\sigma$ by the canonical momentum $\Pi_j^{\sigma, a}= q^{\sigma,a}_j - eA^a$ in a magnetic field $B=\epsilon^{ab}\partial_a A_b$, and thus $R_j^{\sigma, a}$  do not commute. The same density algebra, Eq.~(\ref{commutation}), is also obeyed by the FCIs in the thermodynamic and long-wavelength limit~\cite{ParameswaranRoySondhi,bernevig2012emergent,SOM}, with magnetic field replaced by mean Berry curvature~\cite{SOM}. We will  work in this limit, and will henceforth not distinguish between FQH and FCI.
Our Floquet-FQH approach is based on the observation that the repeated application of the GMP algebra produces $(2N-1)$-body terms from the commutator of two $N$-body terms. In particular, the commutator of two $2$-body terms yields a potentially desirable $3$-body term.

\begin{figure}[t]
 \includegraphics[draft=false,width=\linewidth]{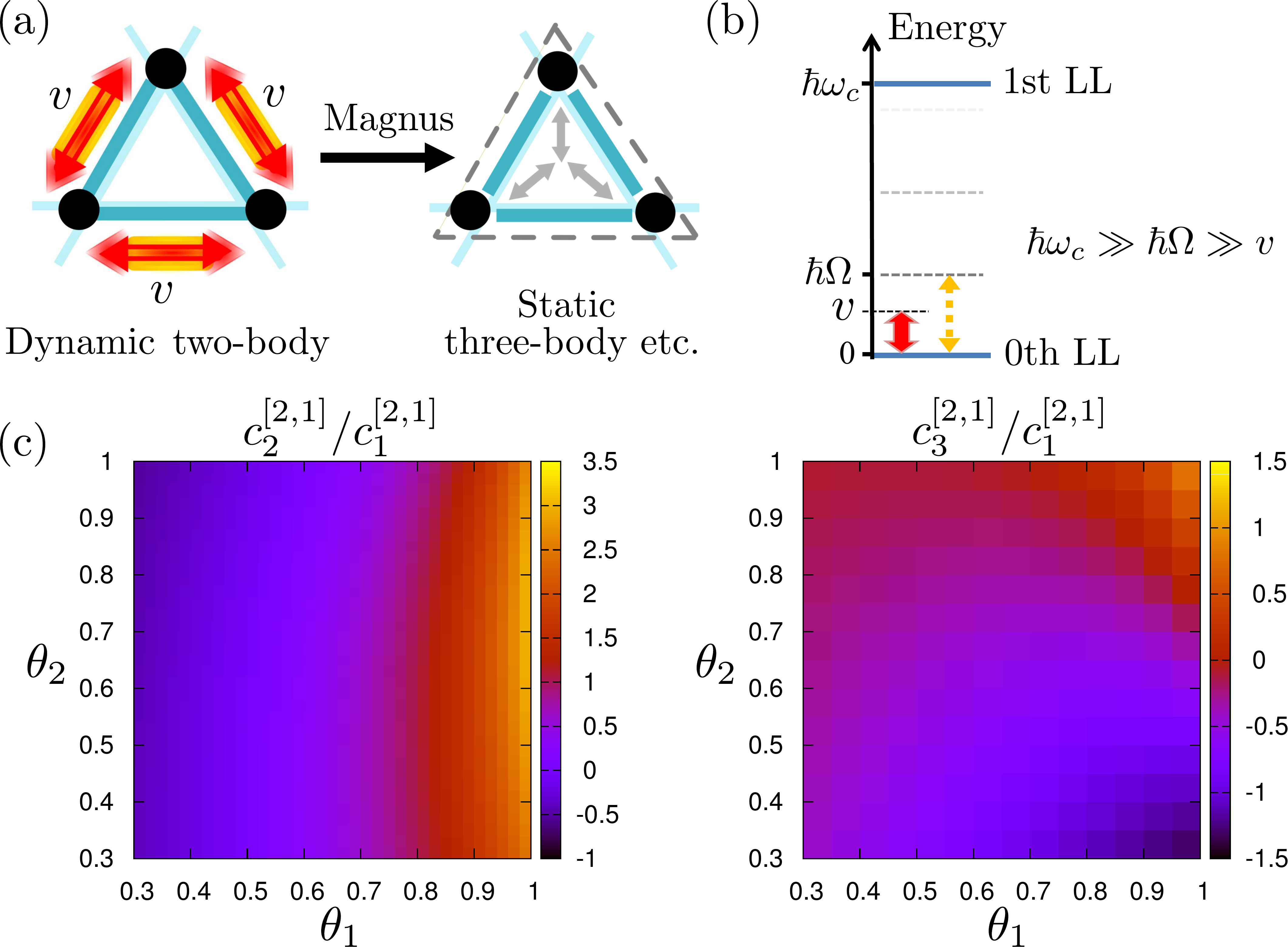}
\caption{ (a) Time-modulated 2-body interactions give rise to an effective 3-body static interaction connecting different sublattices at leading order in $\Omega^{-1}$ (Eq.~\ref{eq:3body}). (b) Energy hierarchy for the setup, with LL (or band) gap $\hbar \omega_c$ much larger than the driving frequency scale $\hbar \Omega$, which should also dominate the interaction $v$. (c) 3-body PP coefficient ratios, $c_2/c_1$ and $c_3/c_1$, for particles with opposite spins ($\lambda=\left[2,1\right]$). The driven two-body interaction is given in  Eq.~(\ref{2to3body}).
}  
\label{Fig1}
\end{figure}

(ii) At high frequencies $\Omega = 2\pi/T$, the stroboscopic dynamics of a periodically-driven system, $H(t)=\sum_le^{il\Omega t}H_l$, can be captured by the static effective Hamiltonian
\begin{eqnarray}
H^\text{eff} = H_0 + \frac1{\hbar \Omega}\sum_{l}\frac1{l}[H_l,H_{-l}]+ \cdots,
\label{eqn:MagnusExp}
\end{eqnarray}
obtained, e.g., from the Magnus or other equivalent high-frequency expansions~\cite{burum1981magnus,blanes2009magnus,bukov2015universal}. Most saliently, Eq.~(\ref{eqn:MagnusExp}) involves commutators which represent the renormalizing  effects of the drive on the interactions. Thus, we see that dynamically modulating a FQH system, combined with the structure of the GMP algebra in (i),  is a natural way to realize higher-body interaction terms,  Fig.~\ref{Fig1}(a). 

{\it Importance of anisotropy.--}A necessary condition for our Floquet-FQH approach is that the commutators in Eq.~(\ref{eqn:MagnusExp}) do not vanish  (this does happen if the system is rotationally symmetric).
Such commutators, however, can be shown to generically survive in \emph{anisotropic} FQH systems. Our protocol is thus targeted at FQH systems with anisotropic interactions; note that this is not a major restriction because anisotropy is ubiquitous in many setups: it can be induced  by tilting the magnetic field~\cite{BoYangPhysRevB.85.165318, PapicTilt,yang2017anisotropic}, and it is intrinsically large in FCIs~\cite{yang2017generalized}.

We remark that the anisotropy of FQH/FCI systems can be quantified using standard Haldane pseudopotentials (PPs) and their generalizations to $N$-particle interactions with internal degrees of freedom~\cite{simon2007,davenport2012multiparticle, lee2013pseudopotential}, which we briefly review now. 
First, one defines a relative angular momentum eigenbasis, $|m\rangle$, in the LL-projected Hilbert space  of $N$ particles with a given permutation symmetry type $\lambda$~\cite{simon2007,davenport2012}. Any isotropic interaction potential $V_\mathbf{q}$ can be expanded in terms of PPs,  $U_{m, \mathbf{q} }^{N,\lambda}$, which form a complete orthonormal basis for $N$-body operators. Below we will use the coefficients in this expansion, $c_m^\lambda$ for fixed $N=3$, in order to characterize the 3-body interactions generated by the Floquet-FQH protocol. The same formalism allows to describe anisotropy by a redefinition of $U_{m, \mathbf{q}}^{N,\lambda} \to U_{m,\Delta m, \pm,  \mathbf{q}}^{N,\lambda}$, where $\Delta m=0,2,4,\ldots$ and $\pm$ denote the discrete symmetry ($U_{m,\Delta m}^{N,\lambda} \propto (q_x + i q_y)^{\Delta m}$) and directionality of the anisotropic PP~\cite{yang2017generalized, SOM}. The coefficients of generalized PPs, $c_{m,\Delta m,\pm}^{N,\lambda}$, completely characterize any translation-invariant interaction and determine which FQH states are energetically favored~\cite{yang2017generalized}. Note that for $\Delta m= 0$, anisotropic PPs reduce to the standard Haldane PPs~\cite{Haldane83,Prange}.

{\it Floquet-FQH system.--}Building on the two key inspirations above, we consider periodically driving an anisotropic FQH/FCI system such that its two-body interaction term is time-modulated while the single-body term remains static, for instance by ultrafast rotation or by appropriate optical driving as detailed later:
\begin{align}
H_{\rm FQH}(t) = H_{\rm nonint} + H_{\rm int}(t).
\label{eqn:drivenFQH}
\end{align}
Here, $H_{\rm nonint}=\frac1{2m}\sum_{i,\sigma}g^{ab}\Pi^\sigma_{a,i}\Pi^\sigma_{b,i}$, and the metric tensor $g^{ab}$ encodes the anisotropy~\cite{HaldaneGeometry,BoYangPhysRevB.85.165318}. We characterize the 2-body interaction $H_{\rm int}(t)$ with its Fourier harmonics $V_l$ and their momentum-space profiles $V^{\sigma\sigma',l}_{\mathbf{q}}$:
\begin{eqnarray}
H_{\rm int}(t)&=&\sum_{l}e^{i\Omega l t/\hbar}V_l =\sum_{\mathbf{q},l,\sigma\sigma'}e^{i\Omega l t/\hbar} V^{\sigma\sigma',l}_{\mathbf{q}}\bar{\rho}_\mathbf{q}^\sigma \bar{\rho}_{-\mathbf{q}}^{\sigma'}.\;\;\;\;\;
\label{Hintt}
\end{eqnarray}

Let us comment on the three relevant energy scales, shown in Fig.~\ref{Fig1}(b), that are behind Eq.~(\ref{eqn:drivenFQH}): (i) the cyclotron frequency $\hbar \omega_c$, set by the single-body term, (ii) the driving frequency $\hbar \Omega$, and (iii) the typical interaction strength $v$, given by the averaged $|V^l_{\mathbf{q}}|$. 
The cyclotron frequency splits the Hilbert space into energetically separated LLs, while the dynamically modulated interaction connects LLs with amplitude $v$, whilst simultaneously allowing energy to be absorbed or emitted in multiples of $\hbar \Omega$. To achieve interesting physics, we consider smooth (strictly low-harmonic) driving obeying the hierarchy $\hbar\omega_c \gg \hbar \Omega \gg v$, i.e., with driving being ``high frequency'' compared to $v$ but not to $\hbar \omega_c$.

The above considerations allow us to derive an effective static description of the system at stroboscopic times, such that there is approximate energy conservation and an effective long-lived ground state~\cite{ho2016quasi}. To see this, note that LL mixing is suppressed due to large LL gaps, high frequency driving and the absence of high order harmonics. Hence we obtain, via a generalized Schrieffer-Wolff transformation, an effective dynamical description of the system  within each LL~\cite{SOM}: 
\begin{align}
H^{\rm LL}_{\rm FQH}(t)=H_{\rm nonint}+\mathcal{P} H_{\rm int}(t) \mathcal{P}\rightarrow \mathcal{P} H_{\rm int}(t) \mathcal{P}
\label{eqn:drivenFQH_LL}
\end{align}
and $H_{\rm nonint}$ drops out as an irrelevant constant. 

We can further employ  Eq.~(\ref{eqn:MagnusExp}) on the effective dynamical Hamiltonian Eq.~(\ref{eqn:drivenFQH_LL}) to  obtain the effective  \emph{static} description $H^\text{eff}$ of the system  within the lowest LL. This description persists up to the exponentially long heating timescale $t_h \sim \frac{\hbar}{v} \exp(\text{const.} \times \Omega/v)$~\cite{abanin2017rigorous,abanin2017effective,mori2016rigorous}, which is estimated to be on the order of years for the example of a cold-atom setup in Fig.~\ref{Fig3} below. 
Assuming a single driving frequency $\Omega$, 
we have 
$H^{\text{eff}} \approx V_0+\frac1{\hbar \Omega}[V_1,V_{-1}]$.
Using Eq.~(\ref{commutation}), after some commutator algrebra~\cite{SOM}, we obtain
$
H^{\text{eff}}  \approx H_{2b} + H_{3b},
$
where the 2-body term $H_{2b}$ 
is the original static profile $V_0$ modified by an operator ordering correction~\cite{SOM}, and the effective 3-body term is
\begin{eqnarray}
\nonumber H_{3b} &=& - \frac{4}{3\hbar \Omega} \sum_{\alpha,\beta,\gamma = \uparrow,\downarrow} \sum_{\mathbf{q},\mathbf{q}'} \text{Im}^{-} \Big\{ 2V^{\beta\alpha *}_{\mathbf{q}} V^{\beta\gamma}_{\mathbf{q}'} 
+V^{\beta\gamma*}_{\mathbf{q}'} V^{\gamma\alpha}_{\mathbf{q}-\mathbf{q}'} \\
\nonumber &+& V^{\alpha\gamma*}_{\mathbf{q}'-\mathbf{q}} V^{\beta\alpha}_{\mathbf{q}} 
+V^{\gamma\beta*}_{\mathbf{q}'-\mathbf{q}} V^{\alpha\gamma}_{\mathbf{q}}  + V^{\gamma\alpha*}_{\mathbf{q}'} V^{\alpha\beta}_{\mathbf{q}-\mathbf{q}'} \Big\}\\
&& \times \sin\frac{\hat{\mathbf{z}} \cdot (\mathbf{q}\times \mathbf{q}')}{2} \bar\rho^\alpha_\mathbf{q}\bar\rho^\beta_{\mathbf{q}'-\mathbf{q}}\bar\rho^\gamma_{-\mathbf{q}'}, \;\;\;\;\;\label{eq:3body}
\end{eqnarray}
where $\text{Im}^{-} \{ f_{\mathbf{q},\mathbf{q}'} \} \equiv (f_{\mathbf{q},\mathbf{q}'} - f^*_{-\mathbf{q},-\mathbf{q}'})/(2i)$ and $\ell_B=1$.

The 3-body interaction in Eq.~(\ref{eq:3body}) is our central result. This interaction emerges from  the products of Fourier components $V^{\sigma\sigma'}_{\mathbf{q}}$  of the original interaction, see Fig.~\ref{Fig1}(a). Due to $\text{Im}^{-}$, Eq.~(\ref{eq:3body}) does not vanish only if $V^{\sigma\sigma'}_{\mathbf{q}}$ (and index permutations) are complex, i.e., only if the system breaks inversion symmetry, and phase differences exist between the modulations of different interaction components. Consequently, $H_{3b}$ is non-zero only in multicomponent anisotropic FQH systems i.e. FCIs with multiatomic unit cells. This peculiar component dependence makes our Floquet approach particularly suited for engineering multicomponent FQH parent Hamiltonians. Finally, we observe that $H^{\text{eff}}$ is not constrained to be repulsive, and could be used to cancel other repulsive interaction terms in the original interaction. 

{\it Illustrative examples.--}We now illustrate the versatility of the Floquet-FQH approach by some examples of interactions and many-body states it could stabilize. First, consider driving a 2-body interaction
$e^{i\Omega t} \sum_{\mathbf{q}} V_{\mathbf{q}} \bar\rho_{\mathbf{q}}^\uparrow \bar\rho_{-\mathbf{q}}^\downarrow + {\rm h.c.}$, which consists of
the simplest anisotropic PPs with $\Delta m = 2$~\cite{yang2017generalized, SOM}:
\begin{equation}
V_{\mathbf{q}} = \cos\theta_1 U_{0,2} + \sin\theta_1\cos\theta_2 U_{1,2} + \sin\theta_1 \sin\theta_2 U_{2,2},
\label{2to3body}
\end{equation}
where $\theta_1$,$\theta_2$ are free parameters that keep the overall interaction strength fixed, while the prefactors of $U_{m,\Delta m}$ can be negative. Eq.~(\ref{2to3body}) produces a range of Floquet 3-body interactions between particles with opposite spins via Eq.~(\ref{eq:3body}). The resulting PP coefficient ratios,  $c^{\left[2,1\right]}_2/c^{\left[2,1\right]}_1$ and $c^{\left[2,1\right]}_3/c^{\left[2,1\right]}_1$,  are shown in Fig.~\ref{Fig1}(c). We see that PP ratios span a wide range, and can become attractive in certain parameter regimes or strongly suppressed, e.g., $U_{1}^{3,\left[2,1\right]}$ and $U_{2}^{3,\left[2,1\right]}$ might be of comparable strength to each other and twice larger than $U_{3}^{3,\left[2,1\right]}$.

Having demonstrated the tunability of 3-body Floquet PPs, we next consider two examples of exotic FQH states that they could naturally stabilize: the interlayer Pfaffian (iPf) state~\cite{ArdonneIPF, BarkeshliIPF} and the $\nu=1$ permanent state (``111-perm") introduced in Ref.~\onlinecite{Moore1991362} (see also Ref.~\onlinecite{green-10thesis}). The iPf state is a gapped state  at filling factor $\nu=2/3$ with non-Abelian Ising anyons, as well as spin-charge separation~\cite{Ardonne-1742-5468-2008-04-P04016, GeraedtsIPF, PetersonIPF, LiuIPF}. By contrast, the 111-perm state is an intriguing gapless state that is governed by a non-unitary conformal field theory~\cite{Moore1991362,ReadPhysRevB.79.245304}, and represents a critical point between the integer quantum Hall ferromagnet and a paramagnet~\cite{ReadRezayiZeroModes}. 

The stability of these FQH states is determined not only by the generated 3-body PPs, which scale as $v^2/\Omega$, but also by original 2-body PPs, which scale as $v$, and   operator ordering corrections to them from the drive (also of the order $v^2/\Omega$)\textcolor{blue}{~\cite{SOM}}. Thus,  if we target a specific state, the original 2-body interaction should be sufficiently ``close" to its model interaction. Many non-Abelian FQH states can be realized in this way, e.g., the ground state of 2-body PPs, $U_1^{\left[1,1\right]}$ and $U_3^{\left[1,1\right]}$, is believed to be in the Moore-Read phase~\cite{PetersonPark}.
In the presence of weak anisotropy, the drive could then further enhance such states by amplifying the 3-body correlations, and thus the robustness of the FQH state. We now illustrate this using exact diagonalizations of continuum FQH systems on the sphere~\cite{SOM}.

\begin{figure}
 \includegraphics[draft=false,width=\linewidth]{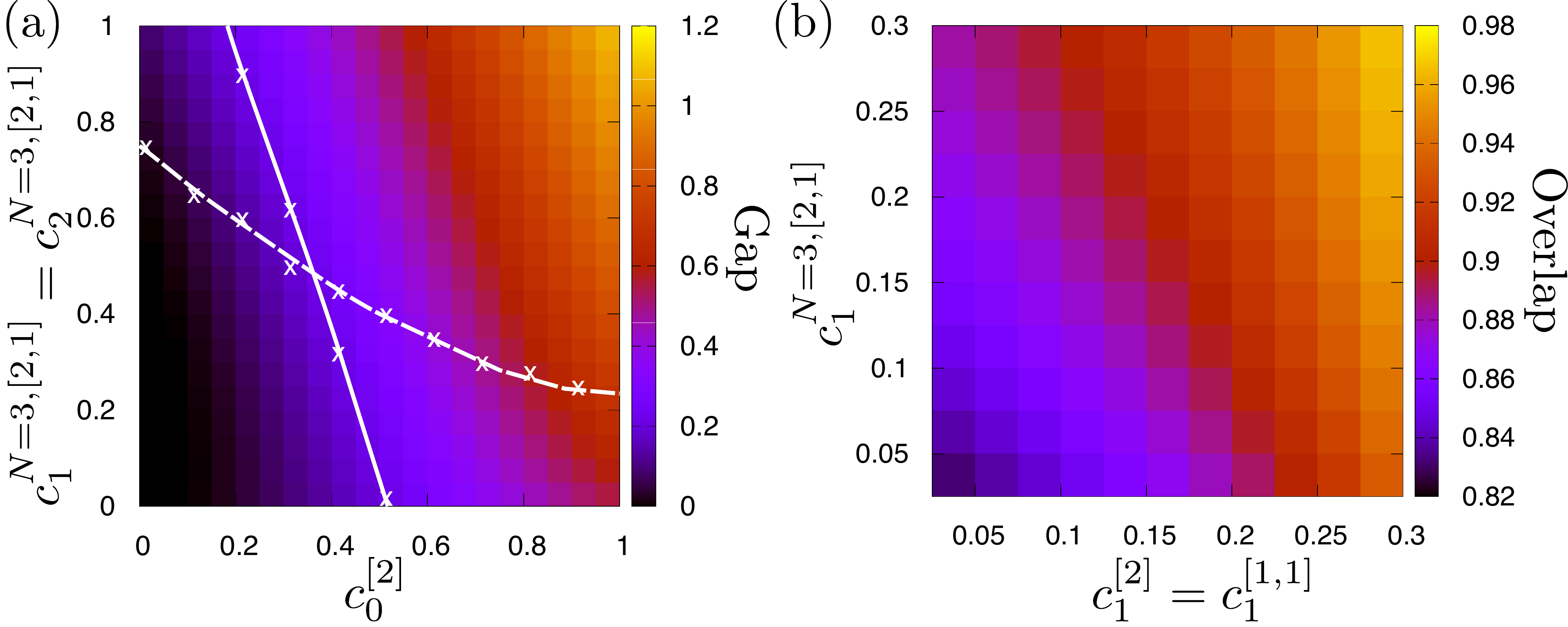}
\caption{
(a) Extrapolated neutral gap (for system sizes $N\leq 10$) as a function of Floquet 2-body PP, $c_0^{\left[2\right]}$, and 3-body PPs, $c_1^{N=3,\left[2,1\right]}= c_2^{N=3,\left[2,1\right]}$. Full line traces gap $\Delta E=0.2$, while dashed line denotes values of the PPs for which the overlap of the ground state and the iPf state (for $N=10$ electrons) is 90\%. The robust iPf phase is stabilized in the top right corner of the phase diagram.
(b) Overlap with 111-perm state is increased by a combination of Floquet 2-body, $c_1^{\left[1,1\right]}=c_1^{\left[2\right]}$,  and 3-body $c_{1}^{N=3,\left[2,1\right]}$ perturbations. Data is for $14$ electrons on the sphere.
}
\label{Fig2main}
\end{figure}

For the iPf we choose the initial ``hollow core'' interaction consisting of 2-body PPs, $U_1^{\left[2\right]}$ and $U_1^{\left[1,1\right]}$, whose strength is fixed to 1. The dominant Floquet corrections  are 2-body $U_0^{\left[2\right]}$, and 3-body $U_1^{N=3,\left[2,1\right]}$ and $U_2^{N=3,\left[2,1\right]}$. In Fig.~\ref{Fig2main}(a) we show the extrapolated neutral gap of the system in the presence of these perturbations.  We assume, for simplicity, that 3-body PPs are of equal magnitude.  The full line in Fig.~\ref{Fig2main}(a) marks the value of the gap $\Delta E = 0.2$, while the dashed line denotes points where the overlap of the ground state and the iPf state is equal to 90\%~\cite{SOM}. Thus, we see that a combination of 2-body and 3-body  Floquet terms results in the large region of a robust iPf phase with non-Abelian correlations and a large gap (top right corner of Fig.~\ref{Fig2main}(a)). 

Similarly, our Floquet approach is also suited for stabilizing the 111-perm state, which crucially relies on a strong $U^{[2,1]}_{1}$~\cite{green-10thesis,CHLeePapicThomale}. In Fig.~\ref{Fig2main}(b), we fix the initial interaction to be $U_0^{\uparrow\downarrow}$ of magnitude 1. The driving is assumed to generate  2-body PPs $c_1^{\left[2\right]}=c_1^{\left[1,1\right]}$ and 3-body PPs $c_1^{N=3,\left[2,1\right]}$, predominantly. By evaluating the overlap with the 111-perm state, we see that the 111-perm phase is enhanced by these perturbations, with the overlap approaching 1. At the same time, the neutral gap of the system remains very small ($\ll 1$) throughout the phase diagram~\cite{SOM}, which is consistent with the gapless phase in the thermodynamic limit~\cite{green-10thesis}. At appropriate filling in bosonic systems, $H^\text{eff}$ with its tail of higher PPs may also stabilize the related 221-permanent state~\cite{ardonne2011,CHLeePapicThomale}.

{\it Experimental proposal.--}In the continuum FQH case,
the Floquet protocol can be implemented by modulating the component of the parallel magnetic field. For magnetic fields $B\sim 20{\rm T}$, this however  requires a very large frequency of $\Omega\sim {\rm 1THz}$.  Instead, a more flexible experimental platform to implement the protocol are FCIs~\cite{neupert2011fractional,sun2011nearly,regnault2011fractional,liu2013review,lee2014lattice,claassen2015position,lee2017band}, which naturally possess large anisotropy, non-trivial unit cell structure and tunable interactions~\cite{yao2012topological,yao2013realizing,peter2015topological,maghrebi2015fractional,yao2015bilayer}.  
We now propose a FCI model of optically driven dipolar spins, realized by trapped dipolar molecules in a 2D optical lattice, which features directional interactions that lead to a direct analogue of $\left[2,1\right]$ 3-body PPs studied above in the continuum FQH case. 

Each molecule in the setup possesses a rovibrational ground state, $|\downarrow\rangle=|0,0\rangle$, and three next-lowest $J=1$ states ($|1,0\rangle$ and $|1,\pm 1\rangle$), which are optically dressed to form a single 'dark' state $|\uparrow\rangle=s|1,-1\rangle + v|1,1\rangle+w|1,0\rangle$, where $s,v$ and $w$ are rational functions of the Rabi frequencies associated with optical driving~\cite{yao2013realizing,SOM}. The $|\uparrow\rangle,|\downarrow\rangle$ states form the effective spin degrees of freedom, which are conserved when the molecules are sufficiently separated such that the physical dipole-dipole interaction between them is much weaker than the bare rotational energy (approximately the Zeeman splitting). In this case, the dipole interaction, together with a strong applied DC field that determines the quantization axis and orbital mixing, is effectively described by hardcore bosons on a lattice with the Hamiltonian~\cite{yao2013realizing}: $H_{\rm FCI}=-\sum_{ij}t_{ij}a^\dagger_ia_j+\frac1{2}\sum_{i\neq j}V_{ij}\rho_i\rho_j$ where $a_i^\dagger=|\uparrow\rangle_i\langle\downarrow|_i$ is the spin-flip operator and $\rho_i=a^\dagger_ia_i$. Both the effective hopping $t_{ij}$ and Hubbard strength $V_{ij}$ originate from the same physical dipole interaction, and can be independently tuned through $\mathbf{E}$ field and the Rabi parameters $s,v,w$ to give rise to FCI states~\cite{yao2013realizing}. 
\begin{figure}[t]
\includegraphics[draft=false,width=\linewidth]{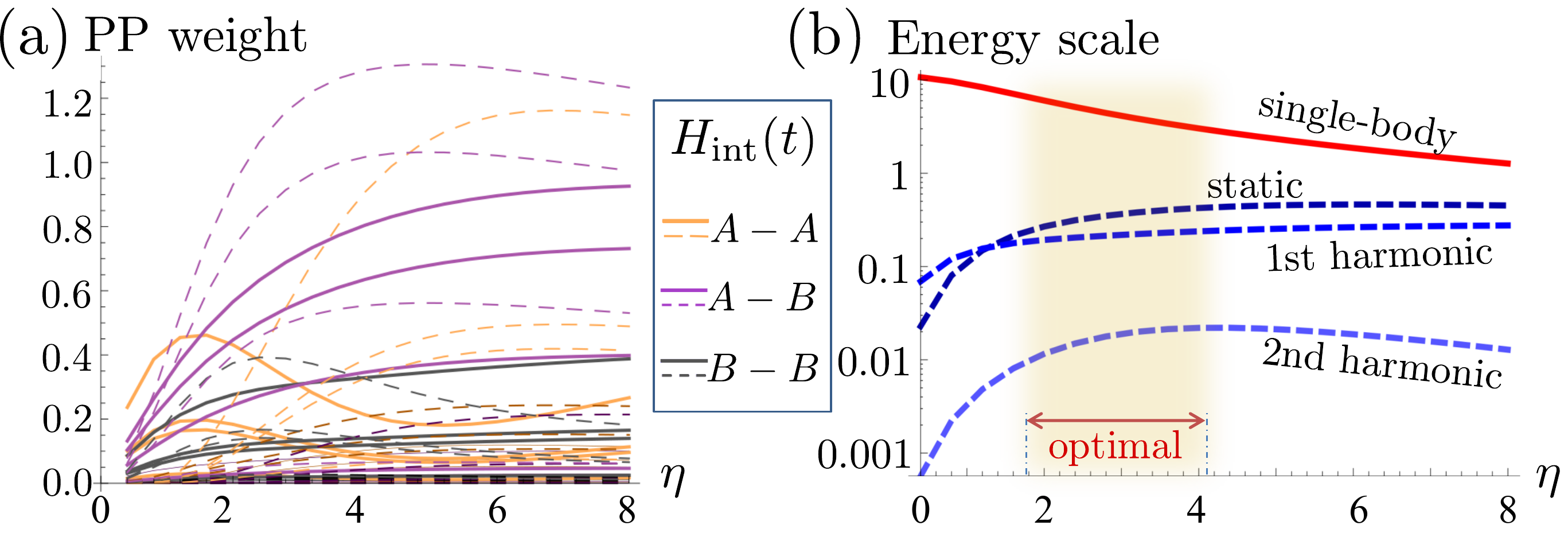}
\caption{(a) Coefficients of 2-body PPs $c_1$, $c_2$, $c_3$ and $c_4$ as a function of $\eta$ for illustrative parameters yielding a band of flatness $\approx 3$~\cite{SOM}. Solid/Dashed curves represent dynamic/static contributions, which are colored according to whether they act between AA, AB or BB sites. (b) Comparison between the energy scales of the single-body, static 2-body ($v$) and lowest two harmonics of the dynamic 2-body parts of $H_{\rm FCI}(t)$ for different $\eta$.  For all purposes, the 2nd harmonic can be neglected. In the optimal shaded regime we have $v\ll \hbar\Omega\ll \hbar\omega_c$.
}
\label{Fig3}
\end{figure}

By modulating the Rabi parameters, it is possible to keep $t_{ij}$ static while $V_{ij}$ is made time-dependent. For an FCI with 2 components $A,B$, we can achieve this by dynamically modulating the Rabi parameters:  
\begin{eqnarray}
\notag s_A(t)=s_A e^{i\Omega_1 t}, \; && \;  s_B(t)=s_B e^{i\Omega_2 t}, \\
\notag v_A(t)=v_A e^{i\Omega_2 t}, \; && \;  v_B(t)=v_B e^{i\Omega_1 t},\\
\notag w_A(t)=W+W'v_Av_B^*e^{-i\Omega t}, \; && \; w_B(t)=W+W's_A^*s_Be^{-i\Omega t},
\end{eqnarray}
where $\Omega=\Omega_2-\Omega_1$ sets the driving frequency, and $W=\sqrt{ \Lambda (1\mp \gamma)/2}$, $W'=\sqrt{\Lambda (1\pm \gamma)/(2v^*_Av_Bs^*_As_B)}$, $\gamma=\sqrt{1- (v^*_Av_Bs^*_As_B/\Lambda^2) (d_{01}/d_{00})^4}$, with $\Lambda$ a real tuning parameter and $d_{01}=\langle 1,\pm 1|d_z|0,0\rangle$, $d_{00}=\langle 1,0|d_z|0,0\rangle$ dipole transition matrix elements that depend on the applied $\mathbf{E}$ field. The Rabi parameter magnitudes are chosen to optimize the band flatness of the resultant tight-binding FCI Hamiltonian~\cite{SOM}, leaving a dynamic 2-body interaction with a single tunable parameter $\eta=2EId/\hbar^2$, the ratio of the molecular dipole energy $Ed$ to its rotational energy scale $\hbar^2/2I$, $I$ being the moment of inertia. 
Coefficients of various 2-body PPs are plotted as a function of \red{$\eta$} in Fig.~\ref{Fig3}(a), and we see that interactions between A and B sites (purple) dominate for most $\eta$. For very small $\eta$, the interaction is mostly dynamical, and its rapid sign fluctuations may destabilize the Floquet ground state. 
The relevant energy scales are shown in Fig.~\ref{Fig3}(b).  In the optimal regime, $1.5\lesssim \eta\lesssim 4$, the single-body hoppings (and hence gap) are one to two orders larger than the interaction, thereby satisfying the requisite hierachy $v\ll \hbar\Omega\ll \hbar\omega_c$. 
At the same time, the static interaction between sublattices is still larger than the dynamic part. Thus, for $\eta\approx 3$ we achieve a direct analog of the above $U_m^{\left[2,1\right]}$ 3-body interaction, assuming we are in  the thermodynamic limit where the GMP algebra is valid. Away from this limit, details of the Bloch wave functions, inter-band transitions and imperfections of the band flatness could affect the stability of the Floquet FCI state.

{\it Conclusions.--}We have presented an approach for generating multicomponent 3-body FQH/FCI interactions, Eq.~(\ref{eq:3body}), via driving anisotropic 2-body interactions with inhomogeneous phase offsets. Our approach is valid in the regime $v\ll \hbar\Omega\ll \hbar\omega_c$, and yields an effective 3-body interaction whose magnitude scales like $\Omega^{-1}$, rather than the conventional $\omega_c^{-1}$ due to LL mixing. The approach relies on the unique properties of the GMP algebra, and thus applies to both FQH and FCI systems in the thermodynamic limit.
We have demonstrated that this approach provides a new route for the exploration of both gapped and gapless multicomponent non-Abelian FQH states, and  proposed its implementation in a Floquet FCI of optically dressed dipolar molecules, where time reversal is broken by the asymmetry between the left and right-circularly polarized optical driving.

{\sl Acknowledgements.--} We thank Nie Wei, Nicolas Regnault and F.D.M. Haldane for helpful discussions. WWH is supported by the Gordon and Betty Moore Foundations EPiQS Initiative through Grant No. GBMF4306. JG is supported by Singapore Ministry of Education Academic Research Fund Tier I (WBS No. R-144-000-353-112) and by the Singapore NRF grant No. NRF-NRFI2017-04 (WBS No. R-
144-000-378-281). ZP acknowledges support by EPSRC grants EP/P009409/1 and EP/R020612/1. Statement of compliance with EPSRC policy framework on research data: This publication is theoretical work that does not require supporting research data.


\bibliography{pseudopotentials}
\clearpage 
\pagebreak

\onecolumngrid
\begin{center}
\textbf{\large Supplemental Online Material for ``Floquet Mechanism for Non-Abelian Fractional Quantum Hall States" }
\end{center}

\vspace{0.1cm}
{\small This supplementary contains the following material arranged by sections:\\
\begin{enumerate}

\item We provide a rigorous treatment of the generalized Schrieffer-Wolff transformation that projects operators into the LLL before they are high-frequency expanded. 

\item We detail and distinguish between the stroboscopic Floquet Hamiltonian and the effective Hamiltonian, and derive the explicit expression for our 3-body effective interaction (Eq. 6 of the main text) via the Magnus expansion. We also provide a detailed discussion and comparison  between the LLL projection in continuum FQH and the band projection in the FCI case.

\item Next we give a pedagogical overview of Haldane pseudopotentials and their generalizations to $N$-body interactions and anisotropic systems.

\item We perform detailed PP decompositions of the effective Floquet interactions considered in the main text. These derivations are supplemented by numerical expressions and further illustrative examples.

\item We provide details of the numerical investigation of FQH states that can be stabilized by 3-body Floquet PPs presented in the main text.

\item Finally, we present details of the dipolar molecule realization of our Floquet approach. We start by reviewing the physical setup, which is then followed by details on how to dynamically modulate the interaction without modulating the single-body part. With that, we discuss our illustrative flatband FCI model and the important role of $\eta$, the tunable ratio between the rotational energy scale and the dipolar energy.
\end{enumerate}
}

\setcounter{equation}{0}
\setcounter{figure}{0}
\setcounter{table}{0}
\setcounter{page}{1}
\setcounter{section}{0}
\makeatletter
\renewcommand{\theequation}{S\arabic{equation}}
\renewcommand{\thefigure}{S\arabic{figure}}
\renewcommand{\thesection}{S\Roman{section}}
\renewcommand{\thepage}{S\arabic{page}}
\vspace{1cm}

\section{I. Derivation of effective Hamiltonians within Landau Levels}
In this section, we derive an effective, {\it time-dependent} Hamiltonian for a periodically driven FQH system that is diagonal within Landau levels (LL) and valid at large cyclotron frequencies $\omega_c \gg \Omega \gg v$ $(\hbar = 1$) for smooth driving protocols. The intuition here is that the large cyclotron gap and low harmonics of the drive suppress  inter-LL processes; thus, this effective Hamiltonian can be thought of as a generalized Schrieffer-Wolff transformation that integrates out such processes.  It is on this effective Hamiltonian, which is diagonal in the LLs, that we can then further employ the Magnus expansion (or other high-frequency expansions) to derive yet another effective Hamiltonian (also within each LL) which is now {\it time-independent}. As discussed in the main text, this is the Hamiltonian  whose effective ground state properties we are interested in.

Let us recall the driven FQH setup. It is of the form
\begin{align}
H_\text{FQH}(t) = H_\text{nonint} + H_\text{int}(t)
\end{align}
with
\begin{align}
H_\text{nonint} & = \frac{1}{2m} \sum_{i, \sigma} g^{ab} \Pi^\sigma_{a,i} \Pi^\sigma_{b,i} = \sum_{i,\sigma} \omega_c (a_i^\sigma)^\dagger a_i^\sigma, \nonumber \\
H_\text{int}(t) &  = \int d^2 \mathbf{q} \sum_{\sigma,\sigma'} V_{\sigma,\sigma'}(\mathbf{q},t) \rho_{\mathbf q}^\sigma \rho_{-\mathbf q}^{\sigma'} \nonumber \\
& =  \int d^2 \mathbf{q} \sum_{\sigma,\sigma'}  V_{\sigma,\sigma'}(\mathbf{q},t) e^{- \frac{1}{2}\mathbf{q}^2} \sum_{i < j} e^{-i\mathbf{q}(\mathbf{R}_i^\sigma - \mathbf{R}_j^{\sigma'})}  \sum_{n_1 n_2 n_3 n_4} 
F_{n_1,n_3}(\bar q) F_{n_2,n_4}(\bar q) V_{i,\sigma}^{n_1 n_3} V_{j,\sigma'}^{n_2 n_4}.
\label{eqn:terms}
\end{align}
In writing the interaction term, we have split the density operators into a part that 
acts within a LL, and a part that moves a particle between LLs (energetically separated by $\omega_c$). The latter is given by bosonic operators $a, a^\dagger$, obtained from diagonalizing the single-particle Hamiltonian, while the former part is defined by the guiding centers $\mathbf{R}_i^\sigma$, which were introduced in the main text~\cite{macdonald1994}. Moreover, we have defined $q\equiv q_x + iq_y$ (with $q^2 = q \bar q = |\mathbf{q}|^2$) and introduced $V_{i,\sigma}^{mn} = (a_{i,\sigma}^\dagger)^m (a_{i,\sigma})^n$. The effective form factors (resulting from scattering between LLs) are given by~\cite{macdonald1994}
$$
F_{n',n}(q) = \sqrt{\frac{n!}{n'!}}\left( \frac{-iq}{\sqrt{2}} \right)^{n'-n} L_n^{n'-n} \left(\frac{q^2}{2} \right).
$$

In what follows, let us assume that the driving $V_{\sigma,\sigma'}(\mathbf{q},t)$ is at frequency $\Omega  = 2\pi/T$ and is smooth, i.e.~contains only strictly low harmonics. We also assume the hierarchy of energy scales that is considered in the main text, $\omega_c \gg \Omega \gg v$. 
We begin by rewriting the Hamiltonian as
\begin{align}
H(t) = \omega_c \left( h_0 + v(t) \right),
\end{align}
where the energy scale $\omega_c$ has been pulled out and  the terms $(h_0, v(t) )$ in the parenthesis correspond to $H_\text{nonint}/\omega_c, H_\text{int}(t)/\omega_c$.  We are interested in the unitary time evolution operator
\begin{align}
U(t) = \mathcal{T} \exp\left(-i \omega_c \int_0^t dt' (h_0 +v(t') ) \right).
\end{align}
In particular, we would like to understand the properties of the Floquet operator $U_F \equiv U(T)$, which is a dynamical map from $t \to t + T$ where $t = T \mathbb{Z}$. 

Consider the following decomposition of $U(t)$ as 
\begin{align}
U(t) = Q(t) \tilde{U}(t) Q^\dagger(0),
\end{align}
where we have yet to define $Q(t)$ aside from the fact that we demand it to be time-periodic and unitary, i.e.~$Q(t) = Q(t + 2\pi/\Omega)$ and $QQ^\dagger = 1$. With this decomposition, at stroboscopic times, $U_F^n = Q(0) \tilde{U}(nT) Q^\dagger(0)$, so the rotation $Q(0)$ can be regarded as static. If it is small (as we will pick it to be in what follows), then for the purposes of measuring local observables, the effect of $Q(0)$ can be ignored -- it is just a small change of frame. Thus, the desired physics is captured solely in $\tilde{U}(t)$.

It is straightforward to check that $\tilde{U}(t)$ obeys the equation of motion
\begin{align}
i \partial_t \tilde{U}(t) = \omega_c {h}'(t) \tilde{U}(t) = \omega_c Q^\dagger(t) \left( h_0 + v(t) - i \frac{\partial_t}{\omega_c} \right) Q(t) \tilde{U}(t), 
\end{align}
therefore
\begin{align}
{h}'(t) \equiv Q^\dagger(t) \left( h_0 + v(t) - i \frac{\partial_t}{\omega_c} \right) Q(t)
\end{align}
 defines a rotated, effective (potentially dynamical) Hamiltonian in this new frame. Since $Q(t)$ is time-periodic, $\tilde{h}(t)$ is as well.

We will choose $Q(t)$ such that the effective Hamiltonian $h'(t)$ is {\it diagonal} in LLs. 
To that end, it will be useful for us to define symmetrization and antisymmetrization operations $\langle . \rangle$ and $\{ . \}$ respectively, which make any operator diagonal or off-diagonal in LLs respectively. 
The symmetrization operator $\langle o \rangle$ on an operator $o$ is defined by 
\begin{align}
\langle o \rangle = \frac{1}{2\pi} \int_0^{2\pi} d\theta e^{i \theta \sum_{i,\sigma} (a_i^\sigma)^\dagger a_i^\sigma } o e^{-i \theta \sum_{i,\sigma} (a_i^\sigma)^\dagger a_i^\sigma },
\end{align}
and the antisymmetrization operator by  $\{ o \} = \langle o \rangle - o$. Thus, we can decompose any operator $o(t)$ (even a time-dependent one) into
\begin{align}
o(t) = \langle o(t) \rangle + \{ o(t) \}.
\end{align}

\subsection{Inverse cyclotron frequency expansion}
Because of the large energy scale $\omega_c$, we are naturally led to consider the expansion of $Q(t)$ as one in powers of the inverse cyclotron frequency, i.e.~
\begin{align}
Q(t) = \exp\left( \sum_{q} S_p(t) \right),
\label{eqn:expansion}
\end{align}
where we implicitly assume that the local norm of $S_p(t) \sim O( \Omega^k v^{p-k} /\omega_c^p)$. Note that this expansion is {\it not} the Magnus expansion, which is a high frequency expansion in the driving frequency $\Omega$. Instead it should be viewed as a generalized Schrieffer-Wolff transformation. 

Using (\ref{eqn:expansion}), 
we can write down the general structure of $h'(t)$ as
\begin{align}
h'(t) & = \sum_{p \geq 0} h^{(p)}(t),
\end{align}
where the local norm of $h^{(p)}(t)   \sim O( (\Omega^k v^{p-k} /\omega_c)^p)$ and
\begin{align}
h^{(0)}(t) & = h_0, \nonumber \\
h^{(1)}(t) & = v(t) - \ad{1} h_0,
\end{align}
and for $ p \geq 2$,
\begin{align}
& h^{(p)}(t) = -\ad{p} h_0 + \left[  (-i) \frac{\partial_t}{\omega_c} S_{p-1}(t) + 
 \sum_{k=2}^p \frac{(-1)^k}{k!} \sum_{\stackrel{1 \leq i_1, \cdots, i_k \leq p}{i_1 + \cdots +i_k = p}} \ad{i_1}\cdots \ad{i_k} h_0  + \right. \nonumber \\
& \left. \sum_{k=1}^{p-1} \frac{(-1)^k}{k!} \sum_{ \stackrel{ 1 \leq i_1, \cdots, i_k \leq p-1}{i_1 + \cdots + i_k = p-1} }  \ad{i_1}  \cdots \ad{i_k} v(t) + 
 i \sum_{m=1}^{p-2} \sum_{k=1}^{p-m-1} \frac{(-1)^{k+1}}{(k+1)!} \sum_{\stackrel{1 \leq i_1, \cdots, i_k \leq p-m-1}{i_1 + \cdots +i_k = p-m-1}} \ad{i_1} \cdots \ad{i_k} \frac{\partial_t}{\omega_c} S_m(t). \right]
 \label{eqn:hp}
\end{align}
where $\ad{p}Y=[S_p,Y]$, $Y$ an arbitrary function. We can rewrite the above as
\begin{align}
h^{(p)}(t) & = -\ad{p}h_0 +  g^{(p)}(t)  \nonumber \\
& = -\ad{p}h_0 +  \{ g^{(p)}(t) \} + \langle g^{(p)}(t) \rangle,
\end{align}
where $g^{(p)}(t)$ is defined to be the term in the square brackets in Eq.~(\ref{eqn:hp}).

Notice that $ g^{(p)}(t)$ is comprised solely of nested commutators of $S_k(t)$ with $h_0$ and $v(t)$ for $k < p$. Thus, we can choose $S_p(t)$ recursively to cancel out LL-transitioning terms at that order, i.e.~we choose $S_p(t)$ such that the following holds:
\begin{align}
 [S_p(t),h_0] = \{ g^{(p)}(t) \}  \text{ for } p \geq 1.
\label{eqn:Sq}
\end{align}
The explicit solution can be written in Fourier space, where $l_p$ corresponds to the Fourier modes:
\begin{align}
 \langle \vec{n} | S_{p,l_p} | \vec{m} \rangle =  \frac{ \langle \vec{n} | \{g^{(p)}_{l_p} \} | \vec{m} \rangle } {\Delta(\vec{n},\vec{m}) }
\end{align}
where $|\vec{n}\rangle = |(n_1, n_2, n_3 \cdots)\rangle$ is the many-particle state corresponding to particle $i$  being in the $n_i$-th LL, and $\Delta(\vec{n}, \vec{m}) = \sum_i |n_i - m_i|$. Necessarily, since $\{ g_{l_p}^{(p)} \}$ is off-diagonal in LL, $\Delta \geq 1$. 

With the relation (\ref{eqn:Sq}), this then defines the resulting effective Hamiltonian
\begin{align}
h'(t) = \sum_{p \geq 0} \langle g^{(p)}(t) 
\label{eqn:heff}
\rangle,
\end{align}
which is periodic in time and diagonal in LLs. Note that  $\langle g^{(0)}(t) \rangle + \langle g^{(1)}(t) \rangle = h_0 + \langle v(t) \rangle$, which is indeed diagonal in LLs. 

\subsection{Validity of expansion}
One might inquire about the validity of the expansion. Essentially, we need that our basic assumption -- that $S_p$ can be organized in inverse powers of the cyclotron freqency, $S_p(t) \sim O( \Omega^k v^{p-k} /\omega_c^p)$ -- is consistent with our solution. In other words, we need to check that the expressions for $S_p(t)$ are asymptotically controlled by the small factor $1/\omega_c$.

If we assume our drive is smooth, i.e.~has only strictly low harmonics to begin with, then this is indeed true. Consider the cleanest case of a harmonic drive such as a pure cosine drive: for example $V_{\sigma, \sigma'}(\mathbf{q}, t) \sim \cos(\Omega t)$, which means that the Fourier harmonics are only $l = \pm 1$. Then it can be seen readily  from Eqns.~(\ref{eqn:hp}), (\ref{eqn:Sq})  that $S_p(t)$ contains harmonics only from $-p$ to $p$, i.e.
\begin{align}
S_p(t) = \sum_{l_p = -p}^p S_{p,l_p} e^{i l_p\Omega t},
\end{align}
which can be shown by induction. Thus, a possible offending term that might invalidate the assumption of the expansion, such as $\frac{\partial_t}{\omega_c} S_{q-1}(t)$ in Eqn.~(\ref{eqn:Sq}) which $\propto \sum_{l_p = -p}^p  \frac{l_p \Omega}{\omega_c} S_{p,l_p} e^{i l_p \Omega t}$, is controlled, because the  factor $l_p$ is always finite for finite $p$.  
Conversely, if we had taken a non-smooth drive (for example a step function, containing infinite harmonics), then we would quickly see that the expansion fails to make sense because $\frac{l_p \Omega}{\omega_c}$ cannot be viewed as `small' if $l_p$ can be infinite. 
Physically, the origin of this phenomenon is simple: if driving were non-smooth, then the driving field can potentially give or take any multiple $m$ of the frequency $\Omega$ so that the large cyclotron gap $\omega_c$ can be made effectively small $\omega_c \to (\omega_c - m \Omega) \ll \omega_c$ so that direct transitions due to the inter-LL mixing terms can occur; conversely, if the driving were smooth, then $m$ is always finite and such direct transitions never occur.

Thus, for smooth driving, the  expansion Eqn.~(\ref{eqn:heff}) can formally be carried out to all orders and the only relevant terms that survive  in the limit $\omega_c \to \infty$ are 
\begin{align}
H_\text{FQH}^\text{LL}(t) & := \omega_c h'(t) \to \omega_c (h_0 + \langle v(t) \rangle) \nonumber \\
& = H_\text{nonint} + \mathcal{P} H_\text{int}(t) \mathcal{P},
\end{align}
as asserted in the main text.

\section{II. Magnus expansion of a time-modulated Hubbard interaction}

Here we detail the Magnus expansion leading to Eq.~6 of the main text. Upon integrating out a period $T$ of the unitary time evolution perator, the (stroboscopic) Floquet Hamiltonian~\cite{bukov2015universal} is expanded order-by-order in $\Omega^{-1}$ as 
\begin{eqnarray}
H_F[t_0]&=&\frac1{T}\int_{t_0}^{T+t_0}H(t)dt + \frac1{2i\hbar T}\int_{t_0}^{T+t_0}dt\int_{t_0}^tdt'[H(t),H(t')]+\text{higher order...}\notag\\
&=&H_0 + \frac1{\hbar \Omega}\sum_l^{\infty}\frac1{l}\left([H_l,H_{-l}]-e^{il\Omega t_0}[H_l,H_0]+e^{-il\Omega t_0}[H_{-l},H_0]\right)+\text{higher order...}\notag\\
&=&H_0 + \frac1{\hbar \Omega}\sum_l^{\infty}\frac1{l}[H_{l}-e^{-il\Omega t_0}H_0,\,H_{-l}-e^{il\Omega t_0}H_0]+\text{higher order...}
\label{magnus1}
\end{eqnarray}
where $H_l$ is the $l$-th Fourier component of the Hamiltonian $H$, and $t_0$ is starting phase of a period (also called the Floquet gauge). We have only displayed the leading nontrivial commutator term in the Magnus expansion, discarding higher order terms proportional to $1/\Omega^2$ or smaller. 

In general, Eq.~(\ref{magnus1}) depends on $t_0$. However, since oscillations are very rapid compared to experimental timescales, it is desirable to consider a \emph{gauge-invariant} version of the stroboscopic Floquet Hamiltonian known as the effective Hamiltonian 
\begin{equation}
H^{\rm eff}= e^{iK[t_0]}H_F[t_0]e^{-iK[t_0]},
\end{equation}
where $e^{iK[t_0]}$ is an unitary rotation via the Kick operator $K[t_0]$, which is defined by $e^{-i\int^{t_2}_{t_1}H(t')dt'}=e^{-iK[t_2]}e^{-iH^{\rm eff}(t_2-t_1)}e^{iK[t_1]}$, $H$ being the original periodic Hamiltonian. In the literature, the combination $e^{-iK[t]}e^{iK[t_0]}$ is also known as the fast-motion unitary operator, which relates the unitary time evolution of the Floquet Hamiltonian with the original periodic Hamiltonian. The effective Hamiltonian has a Magnus expansion~\cite{bukov2015universal}
\begin{eqnarray}
H^{\rm eff}&=&\frac1{T}\int_{0}^{T}H(t)dt + \frac1{2i\hbar T}\int_{0}^{T}dt\int_{0}^tdt'\left(1-\frac{2(t-t')}{T}\right)[H(t),H(t')]+\text{higher order...}\notag\\
&=&H_0 + \frac1{\hbar \Omega}\sum_l^{\infty}\frac1{l}[H_l,H_{-l}]+\text{higher order...}
\label{magnus2}
\end{eqnarray}
which, at leading nontrivial order, can be simply obtained from the stroboscopic Floquet Hamiltonian $H_F[t_0]$ by truncating the terms containing $t_0$. Henceforth, we shall perform our following derivations based on $H^{\rm eff}$, with results for $H_F[t_0]$ at leading order obtainable simply by replacing $H_l\rightarrow H_l-e^{\mp il\Omega t_0}H_0$ [c.f. Eq.~(\ref{magnus1})].

\subsection{FQH Landau level projection}
We now specialize to a FQH system, where we consider a Laudau level (LL) \emph{projected} Hamiltonian for a chosen $l$: $H_l=\sum_{\alpha\beta}\sum_\mathbf{q} V_{\alpha\beta}(\mathbf{q})\bar\rho^\alpha_\mathbf{q}\bar\rho^\beta_{-\mathbf{q}}$. Here we assume that the bare density operators can be simply replaced by their projected versions, as we argued in the previous Section of this supplement starting from the Schrieffer-Wolff transform. Since $\bar\rho^\alpha_\mathbf{q}$ and $\bar\rho^\beta_{-\mathbf{q}}$ always commute, we necessarily have $V_{\alpha\beta,n}(\mathbf{q})=V_{\beta\alpha,n}(-\mathbf{q})$. Furthermore, since FQH interactions should give real energy penalties, $H_{-n}$ must be given by $\sum_\mathbf{q} V^*_{\alpha\beta,-n}(-\mathbf{q})\bar\rho_\mathbf{q}^\alpha\bar\rho_{-\mathbf{q}}^\beta$, i.e. with exponentials of time, but not momentum, being complex conjugated. In an FQH system, the projected density operators obey the GMP algebra (Eq.~1 of the main text), with magnetic length $\ell_B$ set to unity:
\begin{equation}
[\bar \rho_\mathbf{q}^\alpha,\bar \rho_{\mathbf{q}'}^\beta]
=2i\delta^{\alpha\beta}\sin\frac{(\mathbf{q}\times \mathbf{q}')_z}{2}\bar \rho^\alpha_{\mathbf{q}+\mathbf{q}'}.
\label{GMPFQH}
\end{equation} 
This algebra will be slightly modified when we consider an FCI system later. With Eq.~(\ref{GMPFQH}), the leading order Magnus expansion will give rise to a 3-body interaction in $H^{\rm eff}$ as follows: 
\begin{eqnarray}\label{eq:Hl}
&&[H_l,H_{-l}]\notag\\
  &=&  \sum_{\alpha\beta\gamma\delta}\sum_{\mathbf{q},\mathbf{q}'}V_{\alpha\beta}(\mathbf{q})V^*_{\gamma\delta}(-\mathbf{q}') \left[ \bar\rho_\mathbf{q}^\alpha\bar\rho_{-\mathbf{q}}^\beta,\bar\rho_{\mathbf{q}'}^\gamma\bar\rho_{-\mathbf{q}'}^\delta \right]\notag\\
&=& 2i\sum_{\alpha\beta\gamma\delta}\sum_{\mathbf{q},\mathbf{q}'}V_{\alpha\beta}(\mathbf{q})V^*_{\gamma\delta}(-\mathbf{q}')\sin\frac{(\mathbf{q}\times \mathbf{q}')_z}{2}\left( 
-\bar\rho_{\mathbf{q}}^\alpha\bar\rho_{\mathbf{q}'-\mathbf{q}}^\beta\bar\rho_{-\mathbf{q}'}^\delta\delta_{\beta\gamma}   + \bar\rho_{\mathbf{q}}^\alpha\bar\rho_{\mathbf{q}'}^\gamma\bar\rho_{-\mathbf{q}-\mathbf{q}'}^\beta\delta_{\beta\delta}+\bar\rho_{\mathbf{q}+\mathbf{q}'}^\alpha\bar\rho_{-\mathbf{q}'}^\delta\bar\rho_{-\mathbf{q}}^\beta\delta_{\alpha\gamma}-\bar\rho_{\mathbf{q}'}^\gamma\bar\rho_{\mathbf{q}-\mathbf{q}'}^\alpha\bar\rho_{-\mathbf{q}}^\beta\delta_{\alpha\delta}\right)\notag \\
&=&  2i\sum_{\alpha\beta\gamma}\sum_{\mathbf{q},\mathbf{q}'}\sin\frac{(\mathbf{q}\times \mathbf{q}')_z}{2}\left(  V_{\beta\alpha}^*(\mathbf{q})V_{\beta\gamma}(\mathbf{q}')-V_{\alpha\beta}(\mathbf{q})V^*_{\gamma\beta}(\mathbf{q}')+V^*_{\beta\gamma}(\mathbf{q}-\mathbf{q}')V_{\alpha\gamma}(\mathbf{q})-V_{\alpha\gamma}(\mathbf{q}')V_{\alpha\beta}^*(\mathbf{q}'-\mathbf{q}) \right) \bar\rho^\alpha_\mathbf{q}\bar\rho^\beta_{\mathbf{q}'-\mathbf{q}}\bar\rho^\gamma_{-\mathbf{q}'} \notag\\
\end{eqnarray}
such that $H^{\text{eff}}=H_0+\frac1{\hbar l\Omega }[H_l,H_{-l}] = H_{2b} + H_{3b}$ with the emergent three-body contribution $H_{3b}$ [Eq.~6 in main text] given by
\begin{eqnarray}\label{eq:3b}
\notag H_{3b} &=&   -\frac{4}{3\hbar \Omega}\sum_{\alpha\beta\gamma}\sum_{\mathbf{q},\mathbf{q}'} 
 \text{Im}^- \Big\{ 2V_{\beta\alpha}^*(\mathbf{q})V_{\beta\gamma}(\mathbf{q}')+V^*_{\beta\gamma}(\mathbf{q}')V_{\gamma\alpha}(\mathbf{q}-\mathbf{q}')  + V^*_{\alpha\gamma}(\mathbf{q}' - \mathbf{q})V_{\beta\alpha}(\mathbf{q}) \\
 &+& V^*_{\gamma\beta}(\mathbf{q}'-\mathbf{q})V_{\alpha\gamma}(\mathbf{q})+V^*_{\gamma\alpha}(\mathbf{q}')V_{\alpha\beta}(\mathbf{q}-\mathbf{q}')\Big\} 
 \sin\frac{(\mathbf{q}\times \mathbf{q}')_z}{2}  \bar\rho^\alpha_\mathbf{q}\bar\rho^\beta_{\mathbf{q}'-\mathbf{q}}\bar\rho^\gamma_{-\mathbf{q}'},\;\;\;\;\;\;\;
\end{eqnarray}
and the two-body contributions $H_{2b}$ given by the original static 2-body term plus the residual two-body terms:
\begin{eqnarray}\label{eq:2b}
H_{2b} &=& -\frac{2}{\hbar \Omega}\sum_{\alpha\beta}\sum_{\mathbf{q},\mathbf{q}'}\sin^2\frac{(\mathbf{q}\times \mathbf{q}')_z}{2} \Big\{ (V_{\beta\alpha}^*(\mathbf{q})-V_{\alpha\beta}(\mathbf{q}))V_{\beta \beta}(\mathbf{q}')+V_{\beta\beta }(\mathbf{q}-\mathbf{q}')V_{\alpha\beta }(\mathbf{q})-V_{\alpha\beta }(\mathbf{q}')V_{\alpha\beta}^*(\mathbf{q}'-\mathbf{q}) \Big\} \bar\rho_\mathbf{q}^\alpha\bar\rho_{-\mathbf{q}}^\beta\notag\\
&&+\frac{2}{\hbar \Omega}\sum_{\alpha\beta}\sum_{\mathbf{q},\mathbf{q}'}\sin^2\frac{(\mathbf{q}\times \mathbf{q}')_z}{2} \Big\{ V_{\beta\alpha}^*(\mathbf{q}')V_{\alpha\beta}(\mathbf{q})+V_{\beta\alpha}^*(\mathbf{q}')V_{\beta\alpha}(\mathbf{q}'-\mathbf{q})-V_{\alpha\alpha}(\mathbf{q}'-\mathbf{q})(V_{\beta\alpha}^*(\mathbf{q})+V_{\alpha\beta}(\mathbf{q})) \Big\} \bar\rho_\mathbf{q}^\alpha\bar\rho_{-\mathbf{q}}^\beta,\notag\\
\end{eqnarray}
where we have introduced $\text{Im}^- f(\mathbf{q},\mathbf{q}') \equiv  (f(\mathbf{q},\mathbf{q}')-f^*(-\mathbf{q},-\mathbf{q}'))/(2i)$. In deriving Eqs.~(\ref{eq:3b}-\ref{eq:2b}), we have made use of the commutator identity 
\begin{eqnarray}
[AB,CD]=A[B,C]D + AC[B,D] +[A,C]DB + C[A,D]B,
\end{eqnarray}
and have explicitly symmetrized the summand according to the following rule
\begin{eqnarray}
\notag \sum_{\alpha\beta\gamma}\sum_{\mathbf{q},\mathbf{q}'}f_{\alpha\beta\gamma}(\mathbf{q},\mathbf{q}')\bar\rho^\alpha_\mathbf{q}\bar\rho^\beta_{\mathbf{q}'-\mathbf{q}}\bar\rho^\gamma_{-\mathbf{q}'} 
\rightarrow i\sum_{\alpha\beta}\sum_\mathbf{q}\sin\frac{(\mathbf{q}\times \mathbf{q}')_z}{2}\left( f_{\alpha\beta\beta}(\mathbf{q},\mathbf{q}')+f_{\alpha\alpha\beta}(\mathbf{q}',\mathbf{q})-f_{\alpha\beta\alpha}(\mathbf{q}',\mathbf{q}'-\mathbf{q})\right) \bar\rho_\mathbf{q}^\alpha\bar\rho_{-\mathbf{q}}^\beta+\text{3-body}.
\label{commsymm}
\end{eqnarray}
As we can see, the density algebra produces residual 2-body terms that we absorb into the original static 2-body contributions. The nontrivial Magnus expansion contributions to $H^{\rm eff}$ in Eq.~(\ref{eq:3b}) will be simplified for a 2-component system in Sec. IV. Note that since $[H_l,H_{-l}]$ is manifestly invariant under a global phase rotation $H_l\rightarrow e^{i\phi}H_l,H_{-l}\rightarrow e^{-i\phi}H_{-l}$, the effective Hamiltonian is rightly unaffected by a physically irrelevant phase offset of the driving field. To obtain the stroboscopic Floquet Hamiltonian, which \emph{does} depend on a phase offset $t_0$, one simply replaces $V_{\alpha\beta}$ with  $H_l-e^{-il\Omega t_0}H_0$.

\subsection{FCI occupied band projection}

In this section, we discuss  how the Hubbard interaction is Magnus expanded in a FCI system, and outline the connection with LL projection in the continuum FQH case. The FCI, as will be explained later on,  offers a more accessible route to the experimental realization of the Floquet protocol, e.g., in a system of cold dipolar molecules. 

Compared to an FQH system, the main additional complexity in an FCI is the internal structure of the Bloch eigenfunction $u^\alpha_{\mathbf{q}}$, which enters the projection operator $\mathcal{P}$ onto the lowest band via
\begin{equation}
\mathcal{P}=\sum_{\alpha,\beta,\mathbf{q}} u_\mathbf{q}^{\alpha*}u^\beta_{\mathbf{q}}c_\mathbf{q}^{\beta\dagger}|0\rangle\langle 0|c_\mathbf{q}^\alpha =\sum_\mathbf{q}\gamma_\mathbf{q}^{\dagger}|0\rangle\langle 0|\gamma_\mathbf{q},
\end{equation}
where $\gamma_\mathbf{q}=\sum_\alpha u^{\alpha*}_\mathbf{q}c_\mathbf{q}^\alpha$ annihilates a normal mode with $\alpha,\mathbf{k}$ as component and momentum indices, respectively. Projected onto the lowest band by $\mathcal{P}$, the bare density operator $\rho_\mathbf{q}^\alpha=\sum_{\mathbf{k}} c_\mathbf{k}^{\alpha\dagger}c_{\mathbf{q}+\mathbf{k}}^\alpha$ becomes the projected density operator
\begin{eqnarray}
\bar\rho_\mathbf{q}^\alpha=\mathcal{P}\rho_\mathbf{q}^\alpha\mathcal{P}&=& \sum_{\mathbf{q}'}\gamma_{\mathbf{q}'}^{\dagger}|0\rangle\langle 0|\gamma_{\mathbf{q}'}\sum_{\mathbf{k}} c_\mathbf{k}^{\alpha\dagger}c_{\mathbf{q}+\mathbf{k}}^\alpha\sum_{\mathbf{q}''}\gamma_{\mathbf{q}''}^{\dagger}|0\rangle\langle 0|\gamma_{\mathbf{q}''}\notag\\
&=& \sum_{\mathbf{q}',\mathbf{q}''}\gamma_{\mathbf{q}'}^{\dagger}|0\rangle\langle 0|\left[\sum_\mathbf{k}\sum_{\beta\gamma}u^{*\beta}_{\mathbf{q}'}\delta^{\beta\alpha}_{\mathbf{q}',\mathbf{k}}\delta^{\alpha\gamma}_{\mathbf{q}+\mathbf{k},\mathbf{q}''}u^\gamma_{\mathbf{q}''}\right]|0\rangle\langle 0|\gamma_{\mathbf{q}''}\notag\\
&=&\sum_\mathbf{k}u^{\alpha*}_\mathbf{k}u_{\mathbf{k}+\mathbf{q}}^\alpha  \gamma_\mathbf{k}^{\dagger}|0\rangle\langle 0|\gamma_{\mathbf{k}+\mathbf{q}}.
\end{eqnarray}
These projector density operators satisfy~\cite{bernevig2012emergent}
\begin{eqnarray}
[\bar \rho^\alpha_\mathbf{q},\bar \rho^\beta_\mathbf{w}]&=& \sum_\mathbf{k} \left[u^{\alpha *}_\mathbf{k}u^\alpha_{\mathbf{k}+\mathbf{q}}u^{\beta *}_{\mathbf{k}+\mathbf{q}}u^\beta_{\mathbf{k}+\mathbf{w}+\mathbf{q}}-u^{\beta *}_\mathbf{k}u^\beta_{\mathbf{k}+\mathbf{w}}u^{\alpha *}_{\mathbf{k}+\mathbf{w}}u^\alpha_{\mathbf{k}+\mathbf{w}+\mathbf{q}}\right]\gamma_\mathbf{k}^{\dagger}|0\rangle\langle 0|\gamma_{\mathbf{k}+\mathbf{q}+\mathbf{w}}
\end{eqnarray}
In the long wavelength limit, it is well established in the literature that~\cite{ParameswaranRoySondhi,bernevig2012emergent} 
\begin{equation}
u^{\alpha *}_\mathbf{k}u^\alpha_{\mathbf{k}+\mathbf{q}}u^{\beta *}_{\mathbf{k}+\mathbf{q}}u^\beta_{\mathbf{k}+\mathbf{w}+\mathbf{q}}-u^{\beta *}_\mathbf{k}u^\beta_{\mathbf{k}+\mathbf{w}}u^{\alpha *}_{\mathbf{k}+\mathbf{w}}u^\alpha_{\mathbf{k}+\mathbf{w}+\mathbf{q}}\approx \frac{i}{2}(q_iw_j-q_jw_i)F_{ij}(\mathbf{k}),
\end{equation}
where $A_j(\mathbf{k})=-iu^{*\beta}_\mathbf{k}\partial_ju^\beta_\mathbf{k}$ is the gauge connection and $F_{ij}(\mathbf{k})=\partial_iA_j(\mathbf{k})-\partial_jA_i(\mathbf{k})$ is the Berry curvature (Einstein summation is implied). If we further assume that $F_{ij}(\mathbf{k})$ is reasonably uniform in momentum, which can be engineered to arbitrary precision in an FCI model as in Ref.~\cite{lee2017band}, we obtain the FCI density algebra~\footnote{factor of $2$ is due to Einstein summation} 
\begin{eqnarray}
[\bar \rho_\mathbf{q},\bar \rho_\mathbf{w}]&=& \frac{i}{2}(q_iw_j-q_jw_i)\sum_\mathbf{k} F_{ij}(\mathbf{k})\gamma_\mathbf{k}^\dagger|0\rangle\langle 0|\gamma_{\mathbf{k}+\mathbf{q}+\mathbf{w}}\notag\\
&\approx & i(\mathbf{q}\times\mathbf{w})_z \langle F\rangle \bar\rho_{\mathbf{q}+\mathbf{w}}
\end{eqnarray}
where the average Berry curvature $\langle F\rangle=\frac{2\pi C}{(2\pi/a)^2}$ where $C$ is the Chern number and $a$ is the real-space lattice spacing. Comparing with the GMP algebra for FQH systems (Eq.~\ref{GMPFQH}, with magnetic length $\ell_B$ restored), we see that the two algebras are equivalent in the long-wavelength limit if
\begin{equation}
a^2=2\pi \ell_B^2,
\label{GMPFQH2}
\end{equation}
with Chern number taken to be unity for the FQH LLL. To understand the physical significance of Eq.~(\ref{GMPFQH2}), consider the smallest nonzero value for its LHS and RHS. Suppose that the FCI system has $N_x$ by $N_y$ sites, and the (continuum) FQH system has $N_\phi$ flux quanta piercing through it. Then the minimal $\mathbf{q}$ and $\mathbf{w}$ magnitudes are $\frac{2\pi}{N_xa}$ and $\frac{2\pi}{N_ya}$, giving rise to a minimal value of $\frac{2\pi}{N_xN_y}$ on the LHS. Similarly, the minimal RHS value is $\frac{2\pi l^2}{N_\phi l^2}=\frac{2\pi}{N_\phi}$. Comparing, we see that we obtain a long-wavelength limit equivalence of these two systems when $N_\phi=N_xN_y$.

\section{III. Overview of the pseudopotential formalism for isotropic and anisotropic interactions}

\subsection{General overview}

Here, we provide a brief overview of how a generic FQH interaction can be decomposed into standard (two-body) Haldane pseudopotentials~\cite{Haldane1983} (PPs) and their generalizations to $N$-body PPs~\cite{simon2007}. This formalism is then used in Sec. IV to characterize the effective interactions generated by the Floquet protocol, focusing in particular on the resulting 3-body PPs.

FQH states are classified by how fast the wavefunction of small clusters of particles vanishes as particles are brought together to the same point. For example, in the $\nu=\frac1{m}$ Laughlin state, 
\begin{equation}
\langle \{ z \} | m\rangle \propto (z_1-z_2)^m,
\end{equation}
the wave function vanishes as the $m^{th}$-power as two electrons are brought together, where $m$ is also the relative angular momentum quantum number. This can be directly generalized to $N$-body multicomponent interactions, where the basis states take the form $|\vec m,\lambda\rangle$, where $\{\vec m\}$ is a set of $N-1$ independent angular momentum numbers and $\lambda$ indicates the symmetry type corresponding to a specific Young Tableau~\cite{simon2007pseudopotentials,davenport2012multiparticle,CHLeePapicThomale}.

A FQH system has its kinetic energy quenched by the dispersionless LLs, causing particles to ``move'' around each other in peculiar ways due to an inter-particle potential (ordinarily, particles would just fly apart upon repulsion, but in a QH system they are not allowed to accelerate from each other due to the quenched kinetic energy). A great deal of insight into LL physics can be gleaned from the projection of the interaction onto certain chosen sectors of relative angular momentum between small clusters of particles. The coefficients of this projection are the pseudopotential (PP) coefficients. In particular, certain PPs constitute the parent Hamiltonians of well-known FQH states: for instance, the zero-energy ground state of the lowest angular momentum (fermionic) two-body PP is the 1/3 Laughlin state.

The PP coefficients of an $N$-body operator $V$ in a FQH sytem are given by
\begin{equation}\label{eq:c}
c_{(m_1,m_2,m_3,...,m_N)} = \langle m_1,...,m_N|V|m_1,...,m_N\rangle,
\end{equation}
where $|m_1,...,m_N\rangle$ are many-body states with angular momentum numbers $m_1,...,m_N$. By convention, $m_1$ represents the total center of mass angular momentum, which is unimportant for translationally-invariant interactions. One is free to define the $m_j$'s with respect to any sensible (full-rank) set of linear combination of the real-space coordinates of the particles $z_j=x_j-iy_j$. To do so, we first define new coordinates as detailed in the appendix of Ref.~\onlinecite{lee2013pseudopotential}:
\begin{equation}
w_j=\sum_j R_{ij}z_j,
\end{equation}
where $R$ is a rotation matrix ($RR^T=R^TR=\mathbb{I}$). We shall let the angular momentum $m_j$ be conjugate to the coordinate $w_j$. For instance, with 2 bodies we can have $w_1=\frac1{\sqrt{2}}(z_1+z_2)$ and $w_2=\frac1{\sqrt{2}}(z_1-z_2)$, so that $w_1$ and $w_2$ represent the rescaled center-of-mass and relative coordinates, respectively. Since the rotation $R$ is orthonormal, we have the normalization $\sum_j R_{ij}^2=1$.

For 3 bodies, a possible set of new orthonormal coordinates is given by
\begin{align}
w_1&= \frac1{\sqrt{3}}(z_1+z_2+z_3),\notag\\
w_2&=\frac1{\sqrt{2}}(z_1-z_3),\notag\\
w_3&= \frac1{\sqrt{6}}\left((z_1-z_2)+(z_3-z_2)\right),
\label{orthonew}
\end{align} 
where $w_2$ and $w_ 3$ have the interpretations of two-body relative separation and total relative separation respectively. Hence, $m_2$ and $m_3$ correspond to a two-body angular momentum and the total relative angular momentum, respectively.

Since the scalar product $k^T\cdot z$ between momentum and position vectors should remain invariant under a rotation, $k^T\cdot z= k^T\cdot I  \cdot z= (Rk)^T\cdot(Rz)$, i.e. the momentum vector should also transform like $k\rightarrow Rk$. 
From Ref.~\onlinecite{lee2013pseudopotential}, we can hence express the PP coefficients in Eq.~(\ref{eq:c}) as 
\begin{eqnarray}
c_{(m_1,m_2,m_3,...,m_N)}
=(4\pi)^N \left(\prod_j^N\int\frac{d^2k_j}{(2\pi)^2}\right)V(k_1,...,k_N)\prod_j^N \langle m_j| e^{i (Rk)_j\cdot w_j}|m_j\rangle,
\end{eqnarray}
where $V(k_1,...,k_N)$ is the momentum space profile of the interaction. For each $j$, redefining $(Rk)_j\rightarrow k_j$ for notational simplicity, it is well-known that~\cite{macdonald1994}
\begin{eqnarray}
\langle m_j| e^{i k\cdot w_j}|m_j\rangle &=&e^{-|k|^2/2}L_{m_j}\left(|k|^2\right),
\label{mvm}
\end{eqnarray}
where $L_m$ is the $m$th Laguerre polynomial.

For multicomponent (but still isotropic) states labeled by their young Tableaux, their wavefunctions consist of a spatial part multiplied by a component(spin) basis, such that the product as a whole obeys fermionic/bosonic anti/symmetry. The spatial parts themselves, however, only need to obey the partial symmetry dictated by the Young Tableau. Hence the components of the PP coefficients of a translation invariant multicomponent interaction $U$ take the form
\begin{eqnarray}
c_{(m_2\dots m_N)}^{\alpha\beta\gamma...} &=&\langle m_2,...,m_N|U_{\alpha\beta\gamma}|m_2,...,m_N\rangle \nonumber
\\
&=& \prod_j^N\int\frac{d^2k_j}{\pi}e^{-\frac1{2}\left(\sum_jR_{jl}k_l\right)^2}L_{m_j}\left(
\left(\sum_jR_{jl}k_l\right)^2\right)U_{\alpha\beta\gamma...}(k_1,...,k_N)\delta_{\sum_i k_i,  0}
+\text{perms.}
\label{vmm}
\end{eqnarray}
where Greek letters label internal component indices, with the permutations referring to the various possibilities allowed by the Young Tableau. 

\subsection{Example: isotropic 3-body pseudopotentials}

The 3-body effective interaction $V^{N=3}_{q,q'}$ derived in the main text (assumed isotropic for simplicity) can be PP-expanded with 
dominant (isotropic) coefficients using the special case of Eq.~(\ref{vmm}) above with $N=3$:
\begin{equation}
c_{(m_2,m_3)}=\sum_{q,q'} e^{-\frac1{2}Q_2^2 -\frac{1}{2}Q_3^2} L_{m_2}\left(Q_2^2\right)L_{m_3}\left(Q_3^2\right)V^{N=3}_{q,q'},
\end{equation}
where $\vec m=\vec m'=(m_2,m_3)$ are 3-body relative angular momenta, and $Q_j=(Rk)_j$ are their corresponding rotated momenta from $q$,$q'$. 
For the 3-body basis introduced by Eq.~(\ref{orthonew}), we have $(Rk)_1=0$, $(Rk)_2=\frac{k_1-k_3}{\sqrt{2}}$ and $(Rk)_3=\frac{k_1+k_3-2k_2}{\sqrt{6}}\rightarrow  -\sqrt{\frac{3}{2}}k_2$. Exploiting momentum conservation, we rewrite these quantities by relabeling $(k_1,k_2,k_3)$ by all six permutations of the 3-tuple $(q,q'-q,-q')$, yielding $(Rk)_2=\frac1{2}(q+q')^2$, $\frac1{2}(2q'-q)^2$ and $\frac1{2}(2q-q')^2$, and $(Rk)_3=\frac{3}{2}(q-q')^2$, $\frac{3}{2}q^2$ and $\frac{3}{2}q'^2$. Explicitly, the $N=3$ version of Eq.~(\ref{vmm}) is then rewritten as 
\begin{eqnarray}
c_{(m_2,m_3)}^{\alpha\beta\gamma} 
&=& \frac1{\pi^2}\int d^2q \; d^2q'e^{-\frac1{4}(q+q')^2}e^{-\frac{3}{4}(q-q')^2}L_{m_2}\left(\frac1{2}(q+q')^2\right)L_{m_3}\left(\frac{3}{2}(q-q')^2\right)V_{\alpha\beta\gamma}(q,q')+\text{perms.}\notag\\
&=& \frac1{\pi^2}\int d^2q \; d^2q'e^{-(q^2+q'^2-q\cdot q')}L_{m_2}\left(\frac1{2}(q+q')^2\right)L_{m_3}\left(\frac{3}{2}(q-q')^2\right)\times\notag\\
&&\Big\{ (V_{\alpha\beta\gamma}(q,q')+V_{\alpha\beta\gamma}(q'-q,-q)+V_{\alpha\beta\gamma}(-q',q-q'))
+ (V_{\alpha\beta\gamma}(-q',-q)+V_{\alpha\beta\gamma}(q'-q,q')+V_{\alpha\beta\gamma}(q,q-q'))\Big\}. \notag\\
\label{U3}
\end{eqnarray}
This expression will be applied to fully characterize the effective Floquet interactions in Sec. IV below.

\subsection{Generalized pseudopotentials for anisotropic interactions}


In Ref.~\cite{yang2017generalized} it was recently shown that the PP formalism can be extended to characterize any type of translation-invariant interactions, \emph{without} the added assumption of rotational invariance used in the discussion above. Here we briefly review these ``generalized PPs'' which are crucial for the theoretical description of anisotropic FQH systems.

The complete basis of operators, restricting to 2-particle interactions for simplicity, is given by~\cite{yang2017generalized} 
\begin{eqnarray}
U_{m,\Delta m, +} \left(g; \vec q\right) &=& \lambda_{\Delta m} \mathcal{N}_{m,\Delta m} \left(L_m^{\Delta m}\left(|q|^2\right)e^{-\frac{1}{2}|q|^2}\textbf{q}^{\Delta m}+c.c\right),\label{g1s}\\
U_{m,\Delta m,-}\left(g; \vec q\right)&=& -i \mathcal{N}_{m,\Delta m} \left(L_m^{\Delta m}\left(|q|^2\right)e^{-\frac{1}{2}|q|^2}\textbf{q}^{\Delta m}-c.c\right),\label{g2s}
\end{eqnarray} 
where the normalization factors are $\mathcal{N}_{m,\Delta m}\equiv \sqrt{2^{\Delta m-1}m!/(\pi\left(m+\Delta m\right)!)}$, and $\lambda_{\Delta m}=1/\sqrt{2}$ for $\Delta m=0$ or $\lambda_{\Delta m}=1$ for $\Delta m\neq 0$.
As indicated on the LHS, the generalized PP $U_{m,\Delta m,\pm}(g;\vec{q})$ is a function of two integers, $m$ and $\Delta m$, and has directionality ($\pm$). As with standard PPs, integer $m$ is even or odd, depending on the statistics of the particles, while $\Delta m$ takes values $0, 2, 4, \ldots$. For $\Delta m=0$, $m$ is a good quantum number and $U_{m,\Delta m, -}$ identically vanishes, thus we recover the standard Haldane PPs, $U_{m,0,+}$. Other values $\Delta m = 2, 4, \ldots$ correspond to PP which are not fully rotationally invariant, but have a certain discrete symmetry: e.g., $\Delta m=2$ yields the PP whose isocontours in $q$-space have quadrupolar ($C_4$) symmetry , $\Delta m = 4$ corresponds to octupolar symmetry, etc.  

Apart from the dependence of generalized PPs on the momentum $\vec{q}$, they interestingly also possess a metric degree of freedom, $g$, which is unimodular: ${\rm det}g=1$. The metric $g$ enters the definition of momentum  $\mathbf{q}$ on the RHS of Eq.~(\ref{g1s})-(\ref{g2s}).
To see that, it is convenient to rewrite $g$ as
\begin{eqnarray}
g_{ab}=\omega_a^*\omega_b+\omega_a\omega_b^*,
\end{eqnarray}
with $\omega_a$ as the complex vector satisfying $\epsilon^{ab}\omega_a^*\omega_b=i$. The quantity $\mathbf{q}$ is then defined as
\begin{eqnarray}\label{eq:q}
 \textbf{q}=\omega^aq_a, \;\;\; |q|^2=g^{ab}q_aq_b.
\end{eqnarray}
For the special case of $g^{ab}=\mathbbm{1}$, we have $\omega_x=1/\sqrt{2},\omega_y=i/\sqrt{2}$, which gives  $U_{m,\Delta m} \propto (q_x + i q_y)^{\Delta m}$. 

Since they form a basis, different generalized PPs are properly orthonormal:
\begin{eqnarray}
\int d^2q\; U_{m,\Delta m,\sigma=\pm} \left(g; \vec q\right) U_{m',\Delta m',\sigma'} \left(g; \vec q\right)=\delta_{m,m'}\delta_{\Delta m, \Delta m'}\delta_{\sigma,\sigma'}.
\end{eqnarray}
This allows us to decompose any two-body interaction $V_{\vec{q}}$ over the generalized PPs using 
\begin{eqnarray}
V_{\vec{q}} = \sum_{m,\Delta m, \sigma}  c_{m,\Delta m,\sigma} U_{m,\Delta m,\sigma}(\vec{q}),\;\;\;\;\;
c_{m,\Delta m, \sigma} =  \int d^2q \; V_{\vec{q}} U_{m,\Delta m, \sigma} \left(\vec q\right).\label{decompose2as}
\end{eqnarray}
Here we have suppressed the metric dependence of $U_{m,\Delta m,\sigma}$ (and hence its coefficient $c_{m,\Delta m, \sigma}$) which, in principle, is always present for $\Delta m > 0$. Using Eq.~(\ref{decompose2as}) one can characterize any translation-invariant interaction via a small number of  operators, see, e.g., the case of a FQH system in a tilted magnetic field~\cite{yang2017anisotropic}. For present purposes, we will mostly make use of Eqs.~(\ref{g1s})-(\ref{g2s}) as a starting point to specify anisotropic 2-body interactions that will be dynamically modulated in the Floquet protocol.

\section{IV. Characterization of time-modulated and effective Floquet interactions via 3-body pseudopotentials}


In the previous section, we have arrived at Eq.~(\ref{U3}) which gives the PP coefficient for a particular type of 3-body interaction between particles with opposite spins $\alpha, \beta, \gamma \in \{ \uparrow$, $\downarrow\}$. This expression, however, is in terms of internal component indices, rather than the proper symmetry type, $\lambda$. To specialize to a particular symmetry type, we have to symmetrize the expression in Eq.~(\ref{U3}) across the relevant component subsets. 

For example, for the $[2,1]$ symmetry sector, which will turn out to be relevant to our 3-body effective Floquet interaction, we need to symmetrize over two out of three particles at a time. Replacing $V_{\alpha\beta\gamma}$ by our effective interaction from Eq.~(\ref{eq:3b}), the last line of Eq.~(\ref{U3}) can be simplified to:
\begin{eqnarray}
&& -\frac{8}{3\hbar\Omega}\sum_{\alpha\beta\gamma}\sin\frac{(q\times q')_z}{2}  \notag\\
&& \times \Big\{ \text{Im}^{-}[ V_{\beta\alpha}^*(q)V_{\beta\gamma}(q')+ V^*_{\beta\gamma}(q')V_{\gamma\alpha}(q-q')+ V^*_{\alpha\gamma}(q'-q)V_{\beta\alpha}(q)+ V^*_{\gamma\beta}(q'-q)V_{\alpha\gamma}(q)+ V^*_{\gamma\alpha}(q')V_{\alpha\beta}(q-q')\notag\\
&&+ V_{\beta\alpha}^*(q'-q)V_{\beta\gamma}(-q)+ V^*_{\beta\gamma}(-q)V_{\gamma\alpha}(q')+ V^*_{\gamma\alpha}(-q)V_{\alpha\beta}(q')+ V_{\beta\alpha}^*(-q')V_{\beta\gamma}(q-q')]- (q\rightarrow -q',q'\rightarrow -q) \Big\}, \notag\\
\label{comm3}
\end{eqnarray}
where $\text{Im}^{-}[f(q,q')] \equiv [f(q,q')-f^*(-q,-q')]/(2i)$. The above expression, which contains a total of $2^2\cdot 3^2=36$ terms, has the right symmetry over all spin and momentum indices. We manifestly have $c_{(m_2,m_3)}^{\alpha\alpha\beta}=c_{(m_2,m_3)}^{\alpha\beta\alpha}=c_{(m_2,m_3)}^{\beta\alpha\alpha}$, which together define an unique $c_{(m_2,m_3)}^{[2,1]}$.
Note that because the Gaussian and Laguerre polynomial factors in Eq.~(\ref{U3}) possess the symmetries $q\Leftrightarrow q'$, $(q,q')\Leftrightarrow (-q,-q')$ and $(q_x,q'_x)\Leftrightarrow (q_y,q'_y)$, the PP coefficient $c_{(m_2,m_3)}^{[2,1]}$ vanishes if the components of $V(q)$ are all even in $q$. This means that we must have at least \emph{two} components for the 3-body PPs to be nonzero, since a single-component density-density interaction does not break inversion symmetry and is necessarily even in $q$. 

\subsection{Two-component systems with sublattice symmetry}

To fully exploit the simplifying potential of $\lambda$, we now further specialize to two-component systems with sublattice (particle-hole) symmetry, e.g., a bipartite lattice with the components labelling the sites within a unit cell. This is often a physically realistic assumption, unless the lattice is deliberately given a staggered potential.
With this symmetry, the only unique physical 2-body potentials are
\begin{eqnarray}
 V_{12}(q)=V_{21}(-q)=W(q) \;\;\; {\rm and} \;\;\; V_{11}(q)=V_{22}(q)=Y(q)=Y(-q)=Y^*(q).
\end{eqnarray}
In general, we have $V_{\alpha\beta}(q)=V_{\beta\alpha}(-q)$ because the displacement from $\alpha$ to $\beta$ is spatially inverted from that of $\beta$ to $\alpha$.   

The first observation is that the spin(sublattice)-polarized contributions (i.e., $\lambda=\left[1,1,1\right]$) of the effective 3-body potential [Eq.~(\ref{comm3})] must disappear: $V_{111}=V_{222}\propto 2\times\text{Im}^{-}\left[Y(q)Y(q')+Y(q')Y(q-q')+Y(q'-q)Y(q)\right]\bar\rho^\alpha_q\bar\rho^\beta_{q'-q}\bar\rho^\gamma_{-q'}=0$, since $Y(q)$ is real. That leaves only the $[2,1]$ sector to contain non-vanishing PP coefficients:
\begin{eqnarray}
c_{(m_2,m_3)}^{[2,1]} 
&=& -\frac{16}{3\hbar\Omega}\frac1{\pi^2}\int d^2qd^2q'e^{-(q^2+q'^2-q\cdot q')}L_{m_2}\left(\frac1{2}(q+q')^2\right)L_{m_3}\left(\frac{3}{2}(q-q')^2\right)
\sin\frac{(q\times q')_z}{2}\times\notag\\
&&\text{Im}^{-}[ V_{11}^*(q)V_{12}(q')+ V^*_{12}(q')V_{21}(q-q')+ V^*_{12}(q'-q)V_{11}(q)+ V^*_{21}(q'-q)V_{12}(q)+ V^*_{21}(q')V_{11}(q-q')\notag\\
&&+ V_{11}^*(q'-q)V_{12}(-q)+ V^*_{12}(-q)V_{21}(q')+ V^*_{21}(-q)V_{11}(q')+ V_{11}^*(-q')V_{12}(q-q')]\notag\\
&=& -\frac{16}{3\hbar\Omega}\frac1{\pi^2}\int d^2qd^2q'e^{-(q^2+q'^2-q\cdot q')}L_{m_2}\left(\frac1{2}(q+q')^2\right)L_{m_3}\left(\frac{3}{2}(q-q')^2\right)
\sin\frac{(q\times q')_z}{2}\times\notag\\
&&\text{Im}^{-}[ Y^*(q)W(q')+ W^*(q')W(q'-q)+ W^*(q'-q)Y(q)+ W^*(q-q')W(q)+ W^*(-q')Y(q-q')\notag\\
&&+ Y^*(q'-q)W(-q)+ W^*(-q)W(-q')+ W^*(q)Y(q')+ Y^*(-q')W(q-q')]\notag\\
&=& -\frac{16}{3\hbar\Omega}\frac1{\pi^2}\int d^2qd^2q'e^{-(q^2+q'^2-q\cdot q')}L_{m_2}\left(\frac1{2}(q+q')^2\right)L_{m_3}\left(\frac{3}{2}(q-q')^2\right)
\sin\frac{(q\times q')_z}{2}\times\notag\\
&&\text{Im}^{-}[ Y(q)(W(q')-W(q-q')) + Y(q')(W(q-q')-W(-q))+ Y(q'-q)(W(-q)-W(q'))\notag\\
&&+ W^*(-q)W(-q')+W^*(q')W(q'-q)+ W^*(q-q')W(q)]\notag\\
&=& \frac{16}{3\hbar\Omega}\frac1{\pi^2}\int d^2qd^2q'e^{-(q^2+q'^2-q\cdot q')}L_{m_2}\left(\frac1{2}(q+q')^2\right)L_{m_3}\left(\frac{3}{2}(q-q')^2\right)
U_{[2,1]}(q,q'),
\label{U112m2m3}
\end{eqnarray}
where we have introduced the label $U_{[2,1]}(q,q')$ for the system-dependent part of the integrand,
\begin{eqnarray}
U_{[2,1]}(q,q') \equiv \sin\frac{(q\times q')_z}{2} \text{Im}^{-}[ 2Y(q)(W(q-q')-W(q'))+2Y(q'-q)W(q)+W^*(q)(2W(q-q')-W(q'))].\;\;\;\;\;\;\;\;
\end{eqnarray}
From the first to the second line in Eq.~(\ref{U112m2m3}), we have expressed the components $V_{11},V_{12}$ and $V_{22}$ in terms of $W(q)$ and $Y(q)$. Following that, the expression was simplified through the repeated use of the symmetry of the integrand under inversion $(q,q')\Leftrightarrow (-q,-q')$, as well as its antisymmetry under reflection $q\Leftrightarrow q'$.

\subsection{3-body interaction from 2-body PPs}

In order to systematically understand the implications of Eq.~(\ref{U112m2m3}), we now substitute into it a particular form of the modulated 2-body interactions expressed in terms of PPs. With sufficiently many PP terms, we can approximate any interaction respecting magnetic translation symmetry with arbitrary accuracy, thus our discussion will still remain fairly general. 

Consider a driven interaction given by $H_{int}(t)=e^{i\Omega t}\sum_\mathbf{q} W( q)\rho_{-\mathbf{q}}^\uparrow\rho_{\mathbf{q}}^\downarrow+h.c.+\text{static}$, with the intra-component $Y( q)=0$. Including up to $\Delta m=4 $ anisotropy, we have
\begin{eqnarray}
W( q)&=& e^{-q^2/2}\left(W_0(q^2)+q^2 e^{2i\theta}W_2(q^2)+q^4 e^{4i\theta}W_4(q^2)\right)
\label{Wq}
\end{eqnarray}
where $q_x+iq_y=qe^{i\theta}$. The individual terms $W_0$, $W_2$ and $W_4$ will be fully specified in the following section; here it is sufficient to note that they are all functions of $q^2$, and their subscript represents the value of $\Delta m$. In other words, each $W_{\Delta m}$ is some linear combination of generalized PPs $U_{m,\Delta m,\pm}$ with different $m$'s and with fixed $\Delta m$ (the latter is restricted to 0, 2 or 4). The value of $\Delta m>0$, as mentioned above, indicates the $\Delta m$-fold discrete rotational symmetry of the PP.

Note that the ``angular'' part of the PP, $\propto \mathbf{q}^{\Delta m}$, was extracted in front of each $W_{\Delta m}$ in Eq.~(\ref{Wq}). In the main text and in Eq.~(\ref{Wq}) above, we picked a particular orientation of the PP which gives the isotropic metric in $\mathbf{q}$, and consequently the prefactor $(q_x+i q_y)^{\Delta m}$ of each $W_{\Delta m}$. As shown in Eq.~(\ref{orientation}) below, changing this metric only modifies the  results by an overall scalar prefactor. 

As will be evident shortly, only the anisotropic parts $W_2,W_4$ can contribute to the effective 3-body interaction. 
Substituting $W(q)$ in Eq.~(\ref{U112m2m3}), we obtain 
\begin{eqnarray}
&&U_{[2,1]}(q,q')\notag\\
&=&\sin\frac{(q\times q')_z}{2}\text{Im}^{-}[ W^*(q)(2W(q-q')-W(q'))]\notag\\
&=&2\sin\frac{(q\times q')_z}{2}(q\times q')_z \Big\{ e^{-\frac1{2}(q^2+q'^2)}(q\cdot q')( W_2(q^2)W_2(q'^2)+2\lambda^2_4W_4(q^2)W_4(q'^2)(2(q\cdot q')^2-q^2q'^2))+\text{odd in $\Theta$}\notag\\
&&+2e^{-\frac1{2}((q-q')^2+q^2)}(q\cdot (q-q'))( W_2((q-q')^2)W_2(q^2)+2 W_4((q-q')^2)W_4(q^2)((q^2-q'^2-2q\cdot q')q^2+2(q\cdot q')^2)) \Big\} \notag\\
&\approx &(q^2q'^2-(q\cdot q')^2) \Big\{ e^{-\frac1{2}(q^2+q'^2)}(q\cdot q')( W_2(q^2)W_2(q'^2)+2\lambda^2_4W_4(q^2)W_4(q'^2)(2(q\cdot q')^2-q^2q'^2))\notag\\
&&+2e^{-\frac1{2}(p^2+q^2)}(q\cdot p)( W_2(p^2)W_2(q^2)+2 W_4(p^2)W_4(q^2)(2(q\cdot p)^2-q^2p^2)) \Big\} +\text{odd}\notag\\
&= &e^{-\frac1{2}(q^2+q'^2)}(q^2q'^2-(q\cdot q')^2)(q\cdot q')( W_2(q^2)W_2(q'^2)+2 W_4(q^2)W_4(q'^2)(2(q\cdot q')^2-q^2q'^2))+2\times[q'\rightarrow p]+\text{odd}\notag\\
\end{eqnarray}
where $p=q-q'$. Only terms even in $\Theta=\theta-\theta'=\cos^{-1}\frac{q\cdot q'}{q q'}$ have been explicitly shown, since odd terms evaluate to zero upon integration in Eq.~(\ref{U112m2m3}). In the penultimate line, we have made the small $p,q$ approximation in view of the Gaussian suppression factor $e^{-\frac1{2}(p^2+q^2)}$; additional justification for this approximation is that the integral in Eq.~(\ref{U112m2m3}) will introduce a further $e^{-(q'^2+q^2-q\cdot q')}$ suppression. Note that only the \emph{anisotropic} contributions $\Delta m=2,4$ appear in $U_{[2,1]}(q,q')$. 

Substituting the previous expression for $U_{[2,1]}(q,q')$ into Eq.~(\ref{U112m2m3}), we obtain the 3-body PP coefficients 
\begin{eqnarray}
c_{(m_2,m_3)}^{[2,1]} &=&3\times \frac{16}{3\hbar\Omega}\frac1{\pi^2}\int qdq q'dq'\int d\theta d\theta'e^{-\frac{3}{2}(q^2+q'^2)}e^{qq'\cos\Theta }L_{m_2}\left(\frac1{2}(q+q')^2\right)L_{m_3}\left(\frac{3}{2}(q-q')^2\right)\notag\\
&&\times q^3q'^3 \sin^2\Theta \cos\Theta \left( W_2(q^2)W_2(q'^2)+2 W_4(q^2)W_4(q'^2)q^2q'^2\cos(2\Theta) \right)\notag\\
&=&\frac{8\times 2\pi}{\pi^2\hbar\Omega}\int dq dq'q^4q'^4e^{-\frac{3}{2}(q^2+q'^2)}\int d\Theta e^{qq'\cos\Theta }L_{m_2}\left(\frac1{2}(q^2+q'^2+qq'\cos\Theta)\right)L_{m_3}\left(\frac{3}{2}(q^2+q'^2-qq'\cos\Theta)\right)\notag\\
&&\times \sin^2\Theta \cos\Theta \Big( W_2(q^2)W_2(q'^2)+2 W_4(q^2)W_4(q'^2)q^2q'^2\cos(2\Theta) \Big).
\label{Um2m321}
\end{eqnarray}
In deriving this formula, we have multiplied the integrand by $3$ to account for the $[q'\leftrightarrow p]$ terms, which can be put into identical forms by noticing that $q'$ and $p=q-q'$ play identical roles in the Gaussian factor from the LLL projection: $q^2+q'^2-q\cdot q' = q^2+p^2-q\cdot p$. We have also made a coordinate transformation $\int d\theta d\theta' = \frac1{2}\int d(\theta-\theta')d(\theta+\theta')=\frac{2\pi}{2}\int d\Theta$, where $\Theta=\theta-\theta'$. 

Upon fixing $m_2$ and $m_3$, the Laguerre contributions can be expanded into a linear combination of $\cos n\Theta$, $n=0,1,2,...$, and the $\Theta$ dependence in Eq.~(\ref{Um2m321}) can be integrated out via
\begin{equation}
\int_0^{2\pi}e^{qq'\cos\Theta}\cos (n\Theta) \; d\Theta = 2\pi I_n(qq')
\end{equation}
where $I_n$ is the modified Bessel Function of the first kind. 

\subsection{Explicit evaluation of the leading order 3-body PP coefficients}

So far we have obtained a general expression for 3-body PP coefficients, $c_{(m_2,m_3)}^{\left[2,1\right]}$, in Eq.~(\ref{Um2m321}), without specifying the form of the driven 2-body interaction, other than it contains contributions of $\Delta m = 2$ and $\Delta m =4 $ generalized PPs. Now we fully specify $W_2$ and $W_4$ which will allow us to illustrate what values of 3-body PP coefficients, $c_{(m_2,m_3)}^{\left[2,1\right]}$, we can expect from the effective Floquet interaction. 

For illustration, we consider $W_{\Delta m = 2, 4} = \sum_{m=0,1} \eta_{m,\Delta m} U_{m,\Delta m}$, i.e.,
\begin{eqnarray}
W_2(q^2)=\frac1{4\sqrt{\pi}}\left(\eta_{0,2}+ \frac{\eta_{1,2}(q^2-1)}{\sqrt{3}}\right),\\
W_4(q^2)=\frac1{16\sqrt{3\pi}}\left(\eta_{0,4}+ \frac{\eta_{1,4}(q^2-3)}{\sqrt{5}}\right).
\end{eqnarray}
We have omitted the isotropic contribution $W_0(q^2)$, which we showed to be irrelevant.

To demonstrate the calculation of the PPs, we explicitly evaluate the first few PPs $c_{(m_2,m_3)}^{[2,1]}$. 
Plugging $m_2=m_3=0$ into Eq.~(\ref{Um2m321}), we obtain
\begin{eqnarray}
U_{(0,0)}^{[2,1]} &=&\frac{16}{\pi\hbar\Omega}\int dq dq'q^4q'^4e^{-\frac{3}{2}(q^2+q'^2)}\int d\Theta e^{qq'\cos\Theta }\sin^2\Theta \cos\Theta \notag\\
&&\times \left( W_2(q^2)W_2(q'^2)+2 W_4(q^2)W_4(q'^2)q^2q'^2\cos(2\Theta) \right)\notag\\
&=&2\pi\times\frac{16}{\pi\hbar\Omega}\int dq dq'q^4q'^4e^{-\frac{3}{2}(q^2+q'^2)}\times \frac{ I_2(qq')W_2(q^2)W_2(q'^2)+2 I_4(qq')W_4(q^2)W_4(q'^2)q^2q'^2}{qq'}\notag\\
&=&\frac{12 \eta_{0,2}^2+10 \sqrt{3} \eta_{0,2}\eta_{1,2}+7 \eta_{1,2}^2}{384 \pi \hbar\Omega}+\frac{40 \eta_{0,4}^2+12 \sqrt{5} \eta_{0,4}\eta_{1,4}+7 \eta_{1,4}^2}{40960 \pi\hbar\Omega }\notag\\
&\approx& \left(9.94 \eta_{0,2}^2+14.36\eta_{0,2}\eta_{1,2}+5.80\eta_{1,2}^2+0.31 \eta_{0,4}^2 + 0.21 \eta_{0,4}\eta_{1,4}+0.05\eta_{1,4}^2\right)\times \frac{10^{-3}}{\hbar \Omega}.
\label{U00}
\end{eqnarray}
In the last line, we see that $U_{(0,0)}^{[2,1]}$ depends mostly on the $\Delta m=2$ PPs, which can be very large (in fact dominant) in a FCI.

We next illustrate the $m_2=0, m_3=1$ case:
\begin{eqnarray}
U_{(0,1)}^{[2,1]} &=&\frac{16}{\pi\hbar \Omega}\int dq dq'q^4q'^4e^{-\frac{3}{2}(q^2+q'^2)}\int d\Theta e^{qq'\cos\Theta }\sin^2\Theta \cos\Theta \notag\\
&&\times ( W_2(q^2)W_2(q'^2)+2 W_4(q^2)W_4(q'^2)q^2q'^2\cos(2\Theta) )\left(1-\frac{3}{2}(q^2+q'^2-qq'\cos\Theta)\right)\notag\\
&=&\frac{32}{\hbar \Omega}\int dq dq'q^3q'^3e^{-\frac{3}{2}(q^2+q'^2)}\times [( I_2(qq')W_2(q^2)W_2(q'^2)+2 I_4(qq')W_4(q^2)W_4(q'^2)q^2q'^2)\left(1-\frac{3}{2}(q^2+q'^2)\right)\notag\\
&&+\frac{3}{2}( (I_2(qq')+qq'I_3(qq'))W_2(q^2)W_2(q'^2)+2 ((30/(qq')+qq')I_3(qq')-5I_2(qq'))W_4(q^2)W_4(q'^2)q^2q'^2)]\notag\\
&\approx& -\left(31.08 \eta_{0,2}^2+67.48\eta_{0,2}\eta_{1,2}+34.92\eta_{1,2}^2+1.20 \eta_{0,4}^2 + 1.72 \eta_{0,4}\eta_{1,4}+0.53\eta_{1,4}^2\right)\times \frac{10^{-3}}{\hbar \Omega}
\end{eqnarray}

This can be repeated to evaluate higher 3-body PP coefficients. Since $\Delta m=4$ contributions are much smaller than $\Delta m=2$ contributions, we shall not compute them explicitly. Here we quote the results for the first few 3-body PPs for the case where the dynamically modulated part of the 2-body interaction is given by
\begin{equation}
e^{i\Omega t}(\eta_{02}U_{m=0,\Delta m =2} + \eta_{12}U_{m=1,\Delta m=2} +\eta_{22}U_{m=2,\Delta m=2})+h.c.+ \text{isotropic}.
\end{equation}
Note that this interaction is not necessarily positive definite because the weights $\eta_{02},\eta_{12}$ and $\eta_{22}$ can be negative or complex if there are relative phase factors. With this interaction, we obtain the following $m=0$ to $m=6$ 3-body PPs:
\begin{align}
c^{0,0}_{[2,1]}&= 1 |\eta_{02}|^2+1.36465 ~\text{Re}[\eta_{02} \eta_{12}]+0.509642 |\eta_{12}|^2+ 0.883976  ~\text{Re}[\eta_{02} \eta_{22}]  + 0.753326  ~\text{Re}[\eta_{12} \eta_{22}]  + 0.325202 |\eta_{22}|^2  \notag\\ 
c^{0,1}_{[2,1]}&=-2.95455 |\eta_{02}|^2-6.09319 ~\text{Re}[\eta_{02} \eta_{12}]-2.89629 |\eta_{12}|^2-6.12311  ~\text{Re}[\eta_{02} \eta_{22}]  - 5.62644  ~\text{Re}[\eta_{12} \eta_{22}]  - 2.71762 |\eta_{22}|^2  \notag\\ 
c^{0,2}_{[2,1]}&=2.28926 |\eta_{02}|^2+8.8394 ~\text{Re}[\eta_{02} \eta_{12}]+5.95438 |\eta_{12}|^2+13.4943  ~\text{Re}[\eta_{02} \eta_{22}]  + 15.4568  ~\text{Re}[\eta_{12} \eta_{22}]  + 9.10687 |\eta_{22}|^2  \notag\\
c^{0,3}_{[2,1]}&=0.547709 |\eta_{02}|^2-2.93222 ~\text{Re}[\eta_{02} \eta_{12}]-4.54024 |\eta_{12}|^2-10.2783  ~\text{Re}[\eta_{02} \eta_{22}]  - 17.854  ~\text{Re}[\eta_{12} \eta_{22}]  - 13.9943 |\eta_{22}|^2  \notag\\ 
c^{0,4}_{[2,1]}&=-0.946042 |\eta_{02}|^2-3.08902 ~\text{Re}[\eta_{02} \eta_{12}]-0.699415 |\eta_{12}|^2-1.61981  ~\text{Re}[\eta_{02} \eta_{22}]  + 4.14686  ~\text{Re}[\eta_{12} \eta_{22}]  + 7.39186 |\eta_{22}|^2  \notag\\ 
c^{0,5}_{[2,1]}&=-0.0565815 |\eta_{02}|^2+1.54594 ~\text{Re}[\eta_{02} \eta_{12}]+2.21691 |\eta_{12}|^2+4.64247  ~\text{Re}[\eta_{02} \eta_{22}]  + 7.34576  ~\text{Re}[\eta_{12} \eta_{22}]  + 4.16513 |\eta_{22}|^2 \notag\\ 
c^{0,6}_{[2,1]}&=0.0662517 |\eta_{02}|^2+0.502159 ~\text{Re}[\eta_{02} \eta_{12}]-0.135815 |\eta_{12}|^2-0.255486  ~\text{Re}[\eta_{02} \eta_{22}]  - 3.42807  ~\text{Re}[\eta_{12} \eta_{22}]  - 4.99202 |\eta_{22}|^2.
\label{results}
\end{align}
We have normalized the expressions above by setting the coefficient of $|\eta_{02}|^2$ to be equal to 1 in $c^{0,0}_{[2,1]}$. 

One can choose optimal coefficients between the 2-body PP components to reduce or increase certain 3-body PPs. For instance, with somewhat optimized normalized coefficients $(\eta_{02},\eta_{12},\eta_{22})=(0.646,0.621,0.443)$, we obtain
\[{1.6866, -8.63519, 16.7154, -13.3139, 0.226493, 5.6246, -1.82197} \] 
for $c^{0,m}_{[2,1]}$, $m=0,...,6$, for which the $m=2$ component is made largest and repulsive. 

In an actual modulated interaction with various differing phase factors, we will in general necessarily obtain a number of dynamic PP components with relative complex phases between them. From Eq.~(\ref{results}), we also see that, generically, the effective static 3-body interaction so obtained can be tuned to be attractive in some sectors, and repulsive in others. Although we have only explicitly included the first several anisotropic 2-body PPs, typical FCI interactions have rather long tails of anisotropic PPs which can be larger than the isotropic ones, and hence offer additional degrees of freedom for tuning. Ultimately, the ground state depends not just on the 3-body effective interaction, but also the residual static 2-body interactions. Finally, if the driving frequency scale is lowered to the same order as the interactions, the Magnus expansion breaks down and we obtain significant corrections from additional four-body terms and beyond.

\subsection{Effect of orientation of 2-body anisotropy}

One subtle distinction of anisotropic PPs, evident from their definition in Eqs.~(\ref{g1s})-(\ref{g2s}) above, is that there are two independent orthogonally oriented PPs for a given pair of $m$ and $\Delta m$. Here we briefly discuss the effect of their relative coefficients, which controls the angular profile of their linear combination:
\begin{equation}
|v_+U_{m,\Delta m,+}+ v_-U_{m,\Delta m,-}|^2\propto \frac{|v_+|^2+|v_-|^2}{2}+\frac{|v_+|^2-|v_-|^2}{2}\cos (2\Delta m\theta) +\text{Re}(v_+^*v_-)\sin (2\Delta m\theta),
\end{equation}
where we assumed isotropic metric, $qe^{i\theta}=q_x+iq_y$. For concreteness, consider modifying $W(q)$ in Eq.~(\ref{Wq}) to
\begin{equation}
W(q)=e^{-q^2/2}\Big(v_+(q_x^2-q_y^2)W_2(q^2)+v_-(2q_xq_y)W'_2(q^2)\Big),
\end{equation}
retaining only the important $\Delta m=2$ terms. It is not hard to show that \emph{all} results that follow remain qualitatively the same, up to a prefactor of $\text{Im}(v^*_+v_-)$: 
\begin{equation}
U_{[2,1]}(q,q')\rightarrow \text{Im}(v^*_+v_-)U_{[2,1]}(q,q')
\label{orientation}
\end{equation}
and ditto for the PP coefficients. In Eq.~(\ref{Wq}), we chose $v_+=1$ and $v_-=i$, so that $\text{Im}(v^*_+v_-)$ is trivial. Note that despite the seemingly trivial form of this prefactor, tweaking $v_\pm$ can cause the 3-body interaction to transition from repulsive to attractive or vice versa, which would cause all PPs to simultaneously change sign.

\subsection{Example 1: PP coefficients for effective nearest-neighbor interaction on a honeycomb lattice}

We consider a simple toy example of driven density-density interactions on a (gauge-rotated) honeycomb lattice containing only nearest neighbor (NN) interactions:
\begin{eqnarray}
H(t)&=&2\sum_{i,j\in NN}\cos(\Omega t+\Delta \phi_{j}) \rho_i\rho_j \notag\\
&\propto & \sum_{q}[e^{i\phi_3}+ \sum_{j=1,2} e^{i (\phi_j+k_j)}]\bar\rho^1_q\bar\rho^2_{-q} +\sum_{q}[e^{i\phi_3}+ \sum_{j=1,2} e^{i (\phi_j-k_j)}]\bar\rho^2_q\bar\rho^1_{-q} \notag\\
&= & \sum_{q}[W(q)\bar\rho^1_q\bar\rho^2_{-q}+W(-q)\bar\rho^2_q\bar\rho^1_{-q}] 
\label{Hthoney}
\end{eqnarray}
where $\vec \phi=(\phi_1,\phi_2,\phi_3)$ are the phases of the driving modulation on NN bonds $1,2$ and $3$, and $k_1,k_2$ are momentum components projected to the reciprocal lattice vectors. In general, one can write $W(q),W^*(q)=C_{\pm q}(\vec\phi)\pm i S_{\pm q}(\vec\phi)$, where $S_q(\vec\phi),C_q(\vec\phi)$ are respectively odd and even in the various phase delays $\vec \phi$. Specializing further to the case of $\vec\phi=(0,\Phi,0)$ for illustrative purposes, we obtain
\begin{eqnarray}
c_{(m_2,m_3)}^{[2,1]}
\propto \frac{2}{\pi^2}\int d^2qd^2p \; e^{-(q^2+p^2+p\cdot q)}L_{m_2}\left(\frac1{4}(p+2q)^2\right)L_{m_3}\left(\frac{3}{4}p^2\right)\sin\frac{(q\times p)_z}{2}[C_q(\phi)S_p(\phi)-C_p(\phi)S_q(\phi)],\notag\\
\label{U4}
\end{eqnarray}
which is non-zero and can be directly evaluated.

\subsection{Example 2: Vanishing 3-body PPs for lattices with a single atom per unit cell}

Here we explicitly illustrate how the 3-body effective interactions vanish for single-component systems, e.g., a square lattice of driven density-density interactions. In this system, the NN interactions are delayed by a phase of $\phi$ relative to the next nearest neighbor (NNN) interactions:
\begin{eqnarray}
H(t)&=&\frac1{2}\cos\Omega t \sum_{i,j\in NN}\rho_i\rho_j + \frac1{4}\cos(\Omega t + \phi)\sum_{i,j\in NNN}\rho_i\rho_j\notag\\
&=& \sum_{\mathbf{q}} \Big\{ \cos\Omega t\,(\cos q_x + \cos q_y)+ \cos(\Omega t +\phi)\cos q_x \cos q_y \Big\} \rho_\mathbf{q}\rho_{-\mathbf{q}}. 
\label{Ht}
\end{eqnarray}
This gives rise to
\begin{align}\label{eq:Ht1}
H_1(q)=H_{-1}(q)^* = V(q)&=\cos q_x + \cos q_y + e^{i\phi}\cos q_x \cos q_y, 
\end{align}
and all other Fourier components zero. For arbitrary momenta $p,p'$, it is easy to see that $\text{Im}^{-}[V(p)V^*(p')]$ is even in $p$ and $p'$. Thus, substituting Eq.~(\ref{eq:Ht1}) into Eq.~(\ref{U3}), the PP coefficients $c_{(m_2,m_3)}$ can be explicitly shown to vanish for symmetry reasons.

\section{V. Numerical evidence for Floquet-engineered non-Abelian FQH phases}

In the main text, we have argued that the Floquet approach, which results in tunable 2-body and 3-body PP interactions between particles with opposite spins, could be used to stabilize multicomponent non-Abelian FQH states. To illustrate this, we have considered two examples of FQH states with different physical properties: the interlayer Pfaffian state~\cite{ArdonneIPF, BarkeshliIPF} and the 111-permanent state~\cite{Moore1991362, green-10thesis}. Here we provide further details of our numerical simulations via exact diagonalization in the sphere geometry, which support these conclusions.

As emphasized in the main text, we start from a 2-body anisotropic interaction $V_0$ between particles with opposite spins (or other type of internal degree of freedom), whose strength is proportional to $v$. Upon driving, the static Floquet Hamiltonian acquires corrections that can be expressed as 2-body PPs $H_{\rm 2b}$ and 3-body PPs $H_{\rm 3b}$ (of course, higher order interactions appear as well, which we have neglected in this work). All these corrections are small, $H_{\rm 2b}, H_{\rm 3b} \propto v^2/\Omega$, given that we work in the high-frequency limit. Thus, if we want to stabilize a specific FQH state by the Floquet drive, we need $V_0$ to be sufficiently ``close'' to the model Hamiltonian of that state. As we mention in the main text and below, this is certainly possible to achieve for the cases we are interested in (iPf,111-permanent states) and we expect such 2-body ``approximations" to a given FQH state can be found in general. Assuming that such a $V_0$ exists, the immediate question is whether $H_{\rm 2b}$, $H_{\rm 3b}$, viewed as perturbations, could further stabilize the target state. Because we are dealing with a strongly-correlated system, this question is not obvious and needs to be settled by numerical simulations. 

For numerical simulations, we consider a continuum FQH system at a given filling fraction $\nu$ for some finite number of electrons, $N$. The electrons live on the surface of a sphere, whose area is fixed by $N$ and $\nu$. A magnetic monopole, whose strength is proportional to the area (in units of $2\pi$), is placed in the center of the sphere, radiating a magnetic field perpendicular to the electron gas. The Hamiltonian for the electrons is defined as a sum over PPs, and can be represented as a matrix in the basis of Fock states built from the Landau-level orbitals~\cite{Haldane1983}. We explicitly enforce projection to the lowest LL by discarding the scattering terms that exchange electrons between lowest LL and higher LLs. Using angular momentum quantum number, the exponentially large Hamiltonian matrix can be reduced to smaller blocks, which are then diagonalized by the Lanczos algorithm. 

To determine the phase of the system for a given PP interaction, we focus on three quantities: (i) overlap of the exact ground state of the system and the trial wave function, $|\langle \psi_{\rm trial} | \psi_{\rm exact}\rangle|^2$; (ii) entanglement spectrum of $| \psi_{\rm exact}\rangle$; and (iii) the neutral excitation gap. In order to compute the overlap, we require a trial wave function, $|\psi_{\rm trial}\rangle$, such as the iPf or 111-perm wave function. Although first-quantized forms of these wave functions do exist~\cite{ArdonneIPF, ReadRezayiZeroModes}, converting from first- to second-quantized representation is factorially difficult. Instead, we generate $|\psi_{\rm trial}\rangle$ from their parent Hamiltonians, which are also known for these states in PP form. Having obtained $|\psi_{\rm trial}\rangle$ or $|\psi_{\rm exact}\rangle$, we can characterize their underlying physics using the ``entanglement spectrum"~\cite{LiHaldane}. For simplicity, we focus on the entanglement spectrum corresponding to the bipartition of the system into $A$ and $B$, where $A$ contains roughly half of single-particle orbitals (e.g., indexed by $S$, $S-1$, ..., 1, 0 if $2S$ is odd). The entanglement spectrum represents the set of Schmidt (singular) values of the density matrix $\rho_A = {\rm tr}_B|\psi\rangle \langle \psi|$, obtained after tracing the degrees of freedom in the subsystem $B$. The multiplicities of Schmidt values, in particular, provide information about the underlying topological order~\cite{LiHaldane}. Finally, the neutral gap is defined as the energy difference between the ground state and the lowest excited state (with arbitrary value of angular momentum).  

Our main result for the iPf state was summarized in Fig.2(a) in the main text. There, we assumed $V_0$ to consist of 2-body PPs, $U_1^{\left[2\right]}$ and $U_1^{\left[1,1\right]}$, whose strength was fixed to 1. The dominant terms in the Floquet static Hamiltonian were assumed to be the 2-body $U_0^{\left[2\right]}$, and 3-body $U_1^{N=3,\left[2,1\right]}$ and $U_2^{N=3,\left[2,1\right]}$ (for simplicity, we assumed the strengths of the 3-body terms were equal). In Fig.2(a) in the main text, we plotted the neutral gap of the system, obtained by extrapolating the energies for different $N\leq 10$, as a function of $c_0^{\left[2\right]}$ and $c_1^{N=3,\left[2,1\right]} = c_2^{N=3,\left[2,1\right]}$. For the iPf phase, the magnetic monopole strength should be set to $2S=(3/2)N-3$~\cite{ArdonneIPF}. Without the Floquet corrections, the extrapolated gap was found to vanish; turning on both perturbations, we identified a large gapped region in the phase diagram. 

\begin{figure}
 \includegraphics[draft=false,width=\linewidth]{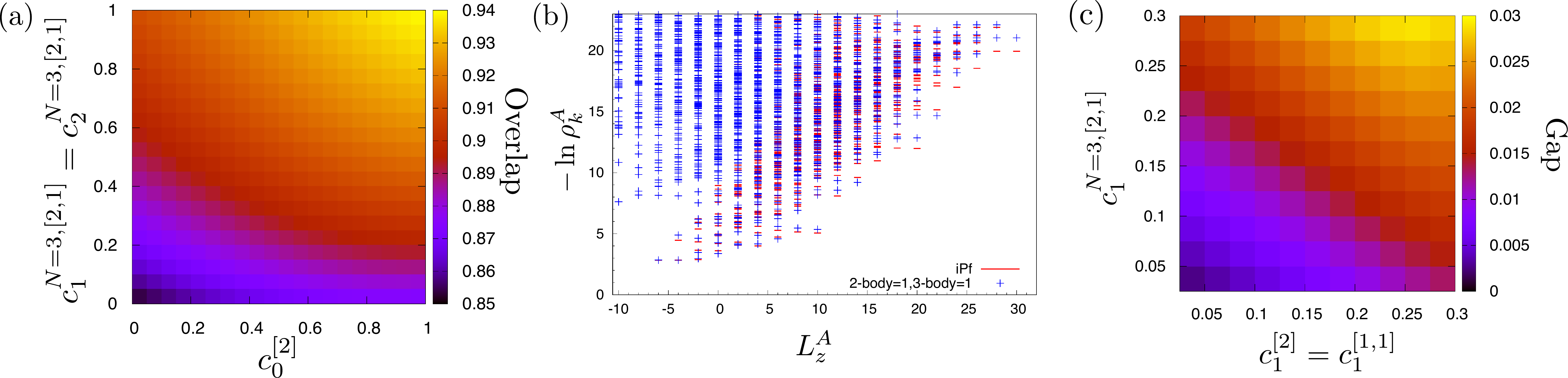}
\caption{(a) Overlap of the iPf state and the ground state of the model interaction which consists of dominant 2-body PPs, $U_1^{\left[2\right]}$ and $U_1^{\left[1,1\right]}$, whose strength is fixed to 1, as a function of perturbations by 2-body $U_0^{\left[2\right]}$, and 3-body $U_1^{N=3,\left[2,1\right]}$ and $U_2^{N=3,\left[2,1\right]}$ (the latter two assumed to be of equal magnitude). The system size is $N=10$ electrons at magnetic flux $2S=12$. 
(b) The low-lying entanglement spectrum of the ground state (blue crosses) for parameters $c_0^{\left[2\right]}= c_1^{N=3,\left[2,1\right]}=c_2^{N=3,\left[2,1\right]}=1$ [top right corner of (a)] matches the exact iPf entanglement spectrum (red line).  Entanglement spectrum is evaluated for the orbital bipartition of the system with 12 electrons and 15 magnetic flux quanta. Subsystem $A$ contains 8 orbitals and 6 electrons (with total spin $S_z^A = 0$). (c) 
Extrapolated neutral gap of the model with dominant 2-body interaction $U_0^{\uparrow\downarrow}$ of magnitude 1, perturbed by 2-body, $c_1^{\left[2\right]}=c_1^{\left[1,1\right]}$, and 3-body $c_1^{N=3,\left[2,1\right]}$ PPs. We consider system sizes $N=8,10,12,14$ with monopole strength $2S=N-2$, corresponding to the 111-perm phase. We see that the gap remains much smaller than any of the PP energy scales. 
}
\label{Fig2}
\end{figure}
In order to confirm that the gapped region corresponds to the iPf phase, we investigate the nature of correlations in the ground state by calculating its overlap with the iPf wave function in Fig.~\ref{Fig2}(a). We see that the overlap remains fairly high (in excess of 80\%) throughout the phase diagram. This is because our choice of $V_0$ is ``nearby" the iPf phase; however, without Floquet perturbations, as we have seen in Fig.2(a) in the main text, the phase is gapless. The high overlaps we see throughout the phase diagram Fig.~\ref{Fig2}(a) should therefore be intepreted with some caution, as the finite-size effects are evidently very strong. As a secondary criterion, which is also well-defined in the thermodynamic limit, we compute the entanglement spectrum in the iPf phase in Fig.~\ref{Fig2}(b). We evaluated the entanglement spectrum by defining the partition $A$ to contain 6 electrons in 8 magnetic orbitals (for the total system of 12 electrons and 16 orbitals). The entanglement spectrum further decomposes into blocks corresponding to the total projection of spin in $A$ (we take the largest block corresponding to $S_z^A=0$). Fig.~\ref{Fig2}(b) shows that the counting of the low-lying entanglement levels (which contain the crucial information about the underlying topological order) is in one-to-one correspondence with the exact iPf entanglement spectrum, suggesting that the ground state is indeed in the iPf phase for the given choice of parameters ($c_0^{\left[2\right]}= c_1^{N=3,\left[2,1\right]}=c_2^{N=3,\left[2,1\right]}=1$).  

Finally, in contrast to gapped topological states, FQH systems can also host intriguing \emph{gapless} phases~\cite{Haldane-PhysRevLett.60.956, ReadRezayiZeroModes, Simon-PhysRevB.75.075317}. These states in many ways resemble their gapped counterparts, e.g., they have elegant bulk wavefunctions which are the ground states of  PP Hamiltonians and obey non-trivial clustering properties~\cite{jack}. However, they can also display unusual properties such as extensive ground state degeneracies~\cite{ReadRezayiZeroModes} or anomalous behavior at the edge of the system~\cite{ReadPhysRevB.79.245304}. A particularly interesting example of such a state is the ``111-permanent" state, introduced in Ref.~\onlinecite{Moore1991362} (see also Ref.~\onlinecite{green-10thesis}). This state is closely related to the $\nu=1$ integer quantum Hall ferromagnet -- it represents a critical state at the transition to a paramagnet~\cite{ReadRezayiZeroModes}. The associated gapless excitations and extensive degeneracy transparently appears in the one-dimensional (``thin torus") limit of a FQH system~\cite{PapicSolvable}.
 
The Floquet-FQH system, where 3-body PP $U_1^{\left[2,1\right]}$ are dominant, may be expected to be a natural host for the 111-permanent state. In the main text, we have computed the overlap of the 111-permanent with the exact ground state of a 2-body interaction  $U_0^{\uparrow\downarrow}$ of fixed magnitude 1, perturbed by 2-body, $c_1^{\left[2\right]}=c_1^{\left[1,1\right]}$, and 3-body $c_1^{N=3,\left[2,1\right]}$ PPs. It was shown that a combination of these those perturbations can lead to very high overlap ($\sim 98\%$) with the 111-permanent state. In Fig.~\ref{Fig2}(c) we show the corresponding neutral gap for the same choice of interactions. The gap was extrapolated from a sequence of system sizes $N=8,10,12,14$ at monopole flux $2S=N-2$. We confirm that in the entire phase diagram the gap is at least an order of magnitude smaller than any of the energy scales of the PPs.

\section{VI. Floquet 3-body interactions between trapped dipolar molecules in an optical lattice}

Here, we detail our approach for generating a 3-body Floquet FQH interaction in a 2D array of trapped dipolar molecules in an optical lattice. Our setup is a dynamical generalization of the effectively static setup first introduced in Ref.~\onlinecite{yao2013realizing}, in which a fractional Chern insulator (FCI) with almost flat Chern bands and 2-body Hubbard-type interactions is realized for certain choices of Rabi parameters. 

\subsection{FCI setup}

Our setup consists of an array of trapped optically dressed dipolar molecules interacting via a physical dipole-dipole interaction $H_{dd}$. Acting on these molecular dipoles is an externally applied electric field $\mathbf{E}$, whose importance as a tuning parameter will be apparent later. The physical Hamiltonian of the molecules is given by
\begin{eqnarray}
H = H_{d}+H_{dd} =  \sum_i\left(\frac{\hbar^2}{2I_i}\hat{\mathbf{J}}_i^2-\mathbf{E}\cdot \mathbf{d}_i\right)+\sum_{i<j}\frac1{4\pi\epsilon_0 |\mathbf{R}_{ij}|^3}\left[\mathbf{d}_i\cdot \mathbf{d}_j -3(\mathbf{d}_i\cdot \mathbf{R}_{ij})(\mathbf{d}_j\cdot \mathbf{R}_{ij})\right],
\label{Hddd}
\end{eqnarray}
where $I_i,\hat{\mathbf{J}}_i^2$ and $\mathbf{d}_i$ are the moment of inertia, squared angular momentum operator and dipole moment of molecule $i$, and $\mathbf{R}_{ij}$ the displacement between molecules $i$ and $j$. Due to the electric field term $\mathbf{E}\cdot \mathbf{d}_i$ in the single-body part of the Hamiltonian $H_d$, the energy eigenstates are no longer exact angular momentum eigenstates. Yet, being adiabatically connected to the true angular momentum eigenstates, they can still be labelled as $|J,M\rangle$, as detailed below. 

\begin{figure}[htb]
\includegraphics[draft=false,width=0.3\linewidth]{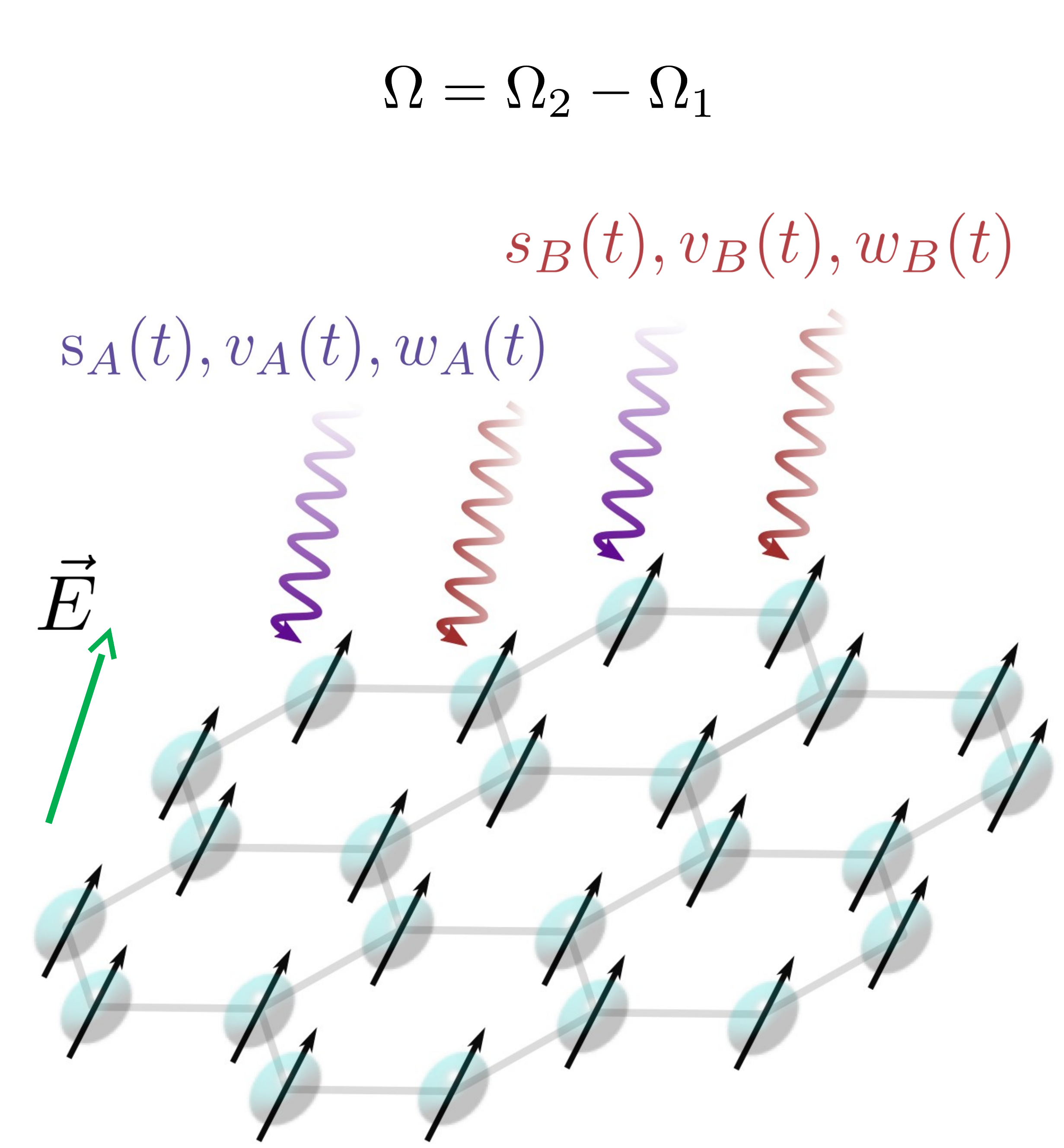}
\caption{A schematic of the driven cold-atom setup for our Floquet FCI. An electrical field $\mathbf{E}$ impinges on the molecular dipoles, which are subject to Rabi parameters that are modulated to produce an effectively driven 2-body interaction at frequency $\Omega=\Omega_2-\Omega_1$.  
}
\label{fig:setup}
\end{figure}

To capture the effect of optical dressing, we focus on the lowest four eigenstates: $|0,0\rangle$,  the rovibrational ground state and $\{ |1,\pm 1\rangle,|1,0\rangle\}$, the $J=1$ multiplet which becomes degenerate when $\mathbf{E}=0$. Optical radiation couples these three $J=1$ states to a pair of excited states in an "M-scheme" described in Ref.~\onlinecite{yao2013realizing}, resulting in an effective two-level system consisting of a "dark" eigenstate, $|\uparrow\rangle=s|1,-1\rangle+v|1,1\rangle + w|1,0\rangle$, and the ground state, $|\downarrow\rangle=|0,0\rangle$, where Rabi parameters $s\propto \Omega_2\Omega_4,v\propto \Omega_1\Omega_4$ and $w\propto \Omega_1\Omega_3$ depend on the four physical Rabi frequencies $\Omega_1,\Omega_2,\Omega_3,\Omega_4$ defining the M-scheme. By expressing the dipole operators in $H_{dd}$ in terms of (pseudo)spin-flip operators $a_i^\dagger=|\uparrow\rangle_i\langle\downarrow|_i$, one can describe optical dressing by an effective 2-component Hamiltonian: 
\begin{equation}
H_{\rm FCI}=-\sum_{ij}t_{ij}a^\dagger_ia_j+\frac1{2}\sum_{i\neq j}V_{ij}\rho_i\rho_j,
\label{HFCIa}
\end{equation}
with $\rho_i=a^\dagger_ia_i$. Both the hopping element, $t_{ij}$, and Hubbard interaction strength, $V_{ij}$, originate from the \emph{same} physical dipole interaction, and can be computed via
\begin{subequations}
\begin{align}
t_{ij}&=\langle \uparrow_i\downarrow_j|H_{dd}|\downarrow_i\uparrow_j\rangle,\label{tijdef}\\
V_{ij}&=\langle \uparrow_i\uparrow_j|H_{dd}|\uparrow_i\uparrow_j\rangle+\langle \downarrow_i\downarrow_j|H_{dd}|\downarrow_i\downarrow_j\rangle-\langle \uparrow_i\downarrow_j|H_{dd}|\uparrow_i\downarrow_j\rangle-\langle \downarrow_i\uparrow_j|H_{dd}|\downarrow_i\uparrow_j\rangle.
\label{Vijdef}
\end{align}
\end{subequations}
The key advantage of this setup is that it is possible to dynamically modulate $V_{ij}$ whilst simultaneously keeping $t_{ij}$ static. This can be achieved via appropriately modulating the Rabi parameters of the optical driving, as we explain below.

We consider an array of dipolar molecules with dipole moments $\mathbf{d}$ whose positions are fixed onto a 2D lattice with two sites per unit cell. In Ref.~\onlinecite{yao2013realizing}, this was chosen to be the checkerboard lattice, which is by no means the only possible choice. The strong applied electric field $\mathbf{E}$ sets the quantization axis of the dipoles. Consider an intermolecular displacement $\mathbf{R} = R(\cos\phi_R,\sin\phi_R,0)= R\hat{\mathbf{R}}$ within the 2D lattice, which we shall align with the $\hat x$-$\hat y$ plane of the lab frame. Letting $\mathbf{E}= E(\sin \lambda\cos\phi_0,\sin\lambda\sin \phi_0,\cos\lambda)$, we see that, in the rotated frame where $\mathbf{E}$ aligns with its $\hat z$ axis, $\hat{\mathbf{R}}$ has the spherical coordinates $(\theta,\phi)$, where $\cos\theta=\sin\lambda\cos(\phi_R-\phi_0)$ and $\sin^2\theta e^{\pm 2i\phi}=(\cos\lambda\cos(\phi_R-\phi_0)\pm i\sin(\phi_R-\phi_0))^2$. These coordinates depend solely on the two angles $\phi_R$ and $\lambda$, which represent the directions of the lattice displacement and the electric field respectively.

\subsection{Rabi parameter modulation that keeps $t_{ij}$ static}

From Eq.~(\ref{tijdef}) or Eqs. S1-S4 of Ref.~\onlinecite{yao2013realizing}, the effective hopping $t_{ij}$ between two molecules separated by $\mathbf{R}=R(\theta,\phi)$ (in spherical coordinates) is given by 
\begin{equation}
t_{ij}=-\frac1{2\pi\epsilon_0 R^3}\left[\left(d_{00}^2w_i^*w_j-\frac1{2}d^2_{01}(v_i^*v_j+s_i^*s_j)\right)C^2_0(\theta)-d_{01}^2\sqrt{\frac{3}{2}}\left(s_i^*v_jC^2_2(\theta,\phi)+v_i^*s_j C^2_{-2}(\theta,\phi)\right)\right]
\label{tij}
\end{equation}
where $v_i,s_i,w_i$ are functions of the Rabi frequency control parameters on site $i$ and 
\begin{align}
 C^2_{\pm 2}(\theta,\phi)&= \sqrt{\frac{15}{32\pi}}\sin^2\theta e^{\pm 2i\phi},\\
C^2_0(\theta) &= \sqrt{\frac{15}{16\pi}}\,(3\cos^2\theta-1), 
\end{align}
are the spherical harmonics with dipolar symmetry, defined w.r.t. to the axis set by $\mathbf{E}$. $d_{00}=\langle 1,0|d_z|0,0\rangle$ and $d_{01}=\langle 1,\pm 1|d_\pm|0,0\rangle$ are molecular dipole transition matrix elements in the basis spanned by $|0,0\rangle$, the rovibrational ground state, and $|J,M\rangle=|1,\pm\rangle,|1,0\rangle$, the lowest excited states of the Hamiltonian $H=H_{dd}-  \mathbf{E}\cdot \mathbf{d} = H_{dd}-Ed_z\cos\theta$.

Importantly, we want to have $t_{ij}$ time-independent while $V_{ij}$ (detailed later) are time-dependent. By inspecting the last term of Eq.~(\ref{tij}), one way this can be implemented is by driving the Rabi variables $s_A,v_A,s_B,v_B$ according to
\begin{align}
s_A & \rightarrow s_A e^{i\Omega_1 t},\\
\label{sA}
v_A & \rightarrow v_A e^{i\Omega_2 t},\\
s_B & \rightarrow s_B e^{i\Omega_2 t},\\
v_B & \rightarrow v_B e^{i\Omega_1 t},
\label{vB}
\end{align}
where $A,B$ label the two sites of the bi-atomic unit cell, and $\Omega_1,\Omega_2$ are two dissimilar driving frequencies. For the left term of Eq.~(\ref{tij}) to be also time-independent, we require
\begin{equation}
w_A^*w_B=\Lambda+D\left(v_A^*v_B e^{i\Omega t}+s_A^*s_B e^{-i\Omega t}\right),
\label{WAB}
\end{equation}
where $\Omega=\Omega_1-\Omega_2$, $D=\frac{d_{01}^2}{2d_{00}^2}$ and $\Lambda$ is a time-independent real tunable parameter. To find the Rabi variables $w_A$ and $w_B$ that satisfy the above, we decompose $w_A^*w_B$ into 
\begin{align}
w_A&=W+W'v_Av_B^*e^{-i\Omega t},\\
w_B&=W+W's_A^*s_Be^{-i\Omega t},
\label{wB}
\end{align}
where $W$ and $W'$ are closed-form functions of $\Lambda$ and $D$ which solve $W^*W'=D$ and $\Lambda=|W|^2+|W'|^2v^*_Av_Bs_A^*s_B$, i.e., $W=\sqrt{\frac{\Lambda}{2}\left(1+ \gamma\right)}$, $W'=\sqrt{\frac{\Lambda}{2v^*_Av_Bs^*_As_B}\left(1- \gamma\right)}$ with $\gamma=\sqrt{1-\frac{v^*_Av_Bs^*_As_B}{\Lambda^2}\left(\frac{d_{01}}{d_{00}}\right)^4}$. With the above time modulations, the hopping term remains static and simplifies to
\begin{equation}
t_{AB}=\sqrt{\frac{15}{4\pi}}\frac{d_{00}^2}{\pi\epsilon_0R^3}\left[\sqrt{3}(s_A^*v_Be^{2i\phi}+v_A^*s_Be^{-2i\phi})\sin^2\theta-\Lambda(3\cos^2\theta-1)\right].
\label{tAB}
\end{equation}
We see that the hoppings $t_{AB}$ are proportional to the square of $d_{00}$, the transition matrix element between $J=1$ and $J=0$ for $M=0$. This is the only factor in $t_{AB}$ that depends on $E$, which thus controls the overall scale of the bandgap (i.e., $\hbar \omega_c$) without modifying the bandstructure. The time-independent part of the Rabi variables $s_A,v_A,s_B,v_B$ shall be chosen to result in almost flat Chern bands just like in Ref.~\onlinecite{yao2013realizing}, except that $w_A,w_B$ can no longer be independently tuned due to the consistency relation, Eq.~(\ref{WAB}). For a specific lattice, doing so also fixes $\Lambda$, but not the relative phase between $s_A$ and $v_B$, nor that between $v_A$ and $s_B$. It is these residual phase freedoms that allow for the possibility of keeping $t_{AB}$ static; in most other systems, any dynamical modulation of the system parameters will invariably modulate the entire effective Hamiltonian. Finally, we mention that there also exist additional on-site inhomogeneities, $t_{ii}$, that can fortunately be regulated via optical tensor shifts~\cite{yao2013realizing}. 

\subsection{Dynamical modulation of the 2-body interaction}

With the form of the time modulation fixed by Eqs.~(\ref{sA})-(\ref{wB}), it is straightforward (though tedious) to obtain the temporal Fourier components of the 2-body Hubbard interaction. From Eq.~(\ref{Vijdef}) and Eqs.~S5 to S10 of Ref.~\onlinecite{yao2013realizing}, the time-dependent Hubbard strengths are given, in terms of the Rabi parameters and various dipole transition matrix elements, by
\begin{eqnarray}
V_{AB}&=&\frac{1}{4\pi\epsilon_0 R^3}\left[\mu^2_{01}\left(\frac1{2}(W_{AB}s_As_B^*+W_{AB}^*v_A^*v_B)e^{i\Omega t}+c.c.\right)-d_{\uparrow_A}d_{\uparrow_B}-d_0^2+d_0(d_{\uparrow_A}+d_{\uparrow_B})\right]C^2_0(\theta)\notag\\
&&+\frac{\mu^2_{01}\sqrt{6}}{4\pi\epsilon_0 R^3}\left[(W_{AB}s_Av_B^*+W_{AB}^*v_A^*s_B)C^2_{-2}(\theta,\phi)+c.c\right]\notag\\
&=&\frac{1}{4\pi\epsilon_0 R^3}\left[\frac{\mu^2_{01}}{2}\left(((\Lambda s_As_B^*+\Lambda^*v_A^*v_B)e^{i\Omega t}+2D(v_A^*v_Bs_As_B^*)e^{2i\Omega t}+D(|s_A|^2|s_B|^2+|v_A|^2|v_B|^2))+c.c.\right)\right]C^2_0(\theta)\notag\\
&&+\frac{1}{4\pi\epsilon_0 R^3}\left[-d_{\uparrow_A}d_{\uparrow_B}-d_0^2+d_0(d_{\uparrow_A}+d_{\uparrow_B})\right]C^2_0(\theta)\notag\\
&&+\sqrt{\frac{45}{32\pi}}\sin^2\theta\frac{\mu^2_{01}}{4\pi\epsilon_0 R^3}\left[(\Lambda s_Av_B^*+\Lambda^*v_A^*s_B+D(s_Av_A^*(|s_B|^2+|v_B|^2)e^{i\Omega t}+s_Bv_B^*(|s_A|^2+|v_A|^2)e^{-i\Omega t}))e^{2i\phi}+c.c\right]\notag\\
\label{VAB}
\end{eqnarray}
where we have introduced the induced dipolar moment $d_{\uparrow_i}=d_1(|s_i|^2+|v_i|^2)+\mu_0|w_i|^2$, $i=A,B$ or, more explicitly,
\begin{align}
d_{\uparrow_A}&=d_1(|s_A|^2+|v_A|^2)+\mu_0(|W|^2+|W'|^2|v_A|^2|v_B|^2+D(v_Av^*_Be^{-i\Omega t}+v_A^*v_Be^{i\Omega t}))=d_A+\mu_0 D(v_Av^*_Be^{-i\Omega t}+v_A^*v_Be^{i\Omega t})\notag\\
d_{\uparrow_B}&=d_1(|s_B|^2+|v_B|^2)+\mu_0(|W|^2+|W'|^2|s_A|^2|s_B|^2+D(s_As^*_Be^{i\Omega t}+s_A^*s_Be^{-i\Omega t}))=d_B+\mu_0 D(s_As^*_Be^{i\Omega t}+s_A^*s_Be^{-i\Omega t})
\label{dAdB}
\end{align}
with $d_A,d_B$ denoting the sum of the static terms on the left. Overall, the Hubbard interaction is proportional to $R^{-3}$, as expected, and depends on the various transition dipole elements $d_{00}=\langle 1,0|d_z|0,0\rangle$, $d_{01}=\langle 1,\pm 1|d_\pm|0,0\rangle$, $\mu_{01}=\langle 1,\pm 1|d_\pm |1,0\rangle$, matrix elements $d_{1}=\langle 1,\pm 1|d_z|1,\pm 1\rangle$, $\mu_0=\langle 1,0|d_z|1,0\rangle$, $d_{0}=\langle 0,0|d_z|0,0\rangle$ as well as Rabi variables $s_A,v_A,s_B,v_B,W,W'$ and auxiliary quantities $D=\frac{d^2_{01}}{2d^2_{00}}$, $\Lambda$, $d_A$, $d_B$.
For subsequent convenience, we also explicitly perform the expansion
\begin{equation}
d_{\uparrow_A}d_{\uparrow_B}=d_Ad_B+\left[\left(D^2s_Av_A^*s_B^*v_Be^{2i\Omega t}+\mu_0 D(d_As_As_B^*+d_Bv_A^*v_B)e^{i\Omega t}+\mu_0^2 s_As_B^*v_Av_B^*\right)+c.c.\right].
\label{dAdB2}
\end{equation}
From Eqs.~(\ref{VAB}), (\ref{dAdB}) and (\ref{dAdB2}), we can collect the static ($n=0$) and dynamic ($n=\pm 1,\pm 2$) Fourier components $V_{AB,n}$ :
\begin{align}
V_{AB,+2}&=\sqrt{\frac{15}{16\pi}}\frac{D}{4\pi\epsilon_0R^3}(v_A^*v_Bs_As_B^*)(\mu_{01}^2-\mu_0^2D)(3\cos^2\theta-1)\notag\\
V_{AB,+1}&=\sqrt{\frac{15}{16\pi}}\frac{1}{4\pi\epsilon_0 R^3}\left[\frac{\mu^2_{01}}{2}\left(\Lambda s_As_B^*+\Lambda^*v_A^*v_B\right)-\mu_0D\left((d_As_As_B^*+d_Bv_A^*v_B)-d_0(v^*_Av_B+s_As_B^*)\right)\right](3\cos^2\theta-1)\notag\\
&+\sqrt{\frac{45}{32\pi}}\sin^2\theta\frac{\mu^2_{01}D}{4\pi\epsilon_0 R^3}\left(s_Av_A^*(|s_B|^2+|v_B|^2)e^{2i\phi}+s_B^*v_B(|s_A|^2+|v_A|^2)e^{-2i\phi}\right)\notag\\
V_{AB,0}&=\sqrt{\frac{15}{16\pi}}\frac{1}{4\pi\epsilon_0 R^3}\left[\mu^2_{01}D\left(|s_A|^2|s_B|^2+|v_A|^2|v_B|^2\right)-(d_A-d_0)(d_B-d_0)-\mu_0^2(s_As_B^*v_Av_B^*+c.c.)\right](3\cos^2\theta-1)\notag\\
&+\sqrt{\frac{45}{32\pi}}\sin^2\theta\frac{\mu^2_{01}}{4\pi\epsilon_0 R^3}\left[(\Lambda s_Av_B^*+\Lambda^*v_A^*s_B)e^{2i\phi}+c.c\right]\notag\\
V_{AB,-1}&=V_{AB,+1}^*\notag\\
V_{AB,-2}&=V_{AB,+2}^*
\label{VABn}
\end{align}
The directional profile of these interaction terms are determined by $\cos\theta=\sin\lambda\cos(\phi_R-\phi_0)$ and $\sin^2\theta e^{\pm 2i\phi}=(\cos\lambda\cos(\phi_R-\phi_0)\pm i\sin(\phi_R-\phi_0))^2$, where $\phi_R$ is the physical angular direction of the interparticle separation (in 2D), and $\phi_0$, $\lambda$ are the angular coordinates of the electric field direction. In previous sections it was shown that the terms with dipole directional dependence are precisely the leading contributors to the Floquet 3-body interaction.

\subsection{Dependence of dipole matrix elements on electric field}

We now discuss the specific dependence of various dipole moments $\langle J',M'|d_z|J,M\rangle$ and transition dipole moments $\langle J',M\pm 1|d_\pm|J,M\rangle$ on the applied electric field strength $E$. The latter constitutes the only free parameter for tuning the interaction independently of the tight-binding bandstructure. 

\begin{figure}
\includegraphics[draft=false,width=0.35\linewidth]{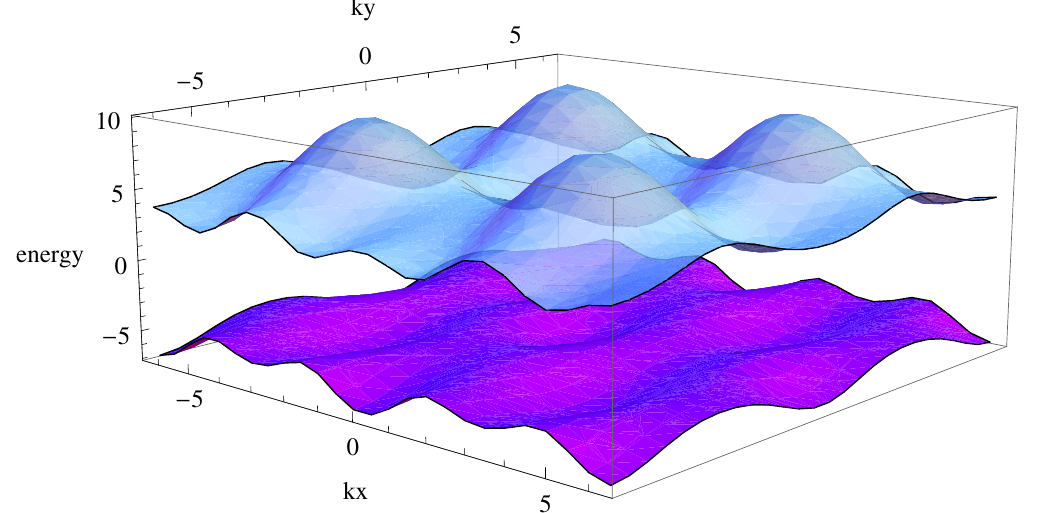}
\includegraphics[draft=false,width=0.32\linewidth]{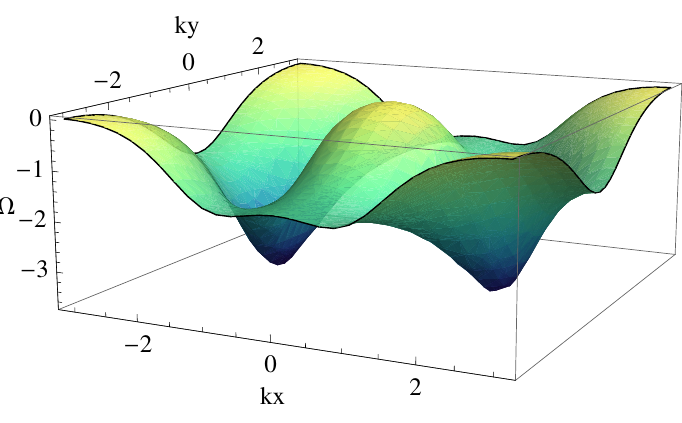}
\includegraphics[draft=false,width=0.31\linewidth]{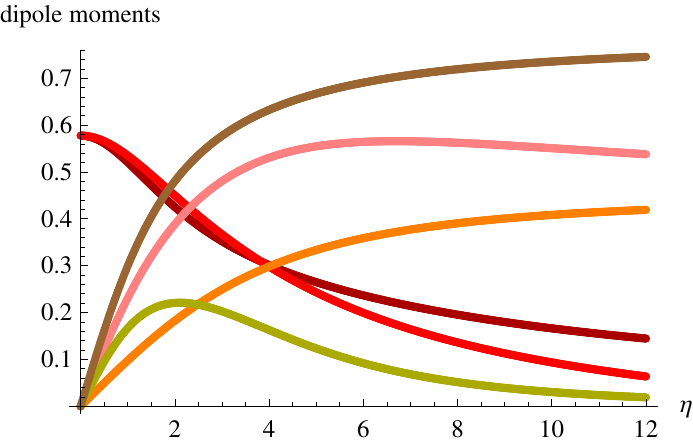}
\caption{ (left) Bandstructure of the flatband Hamiltonian given by Eq.~(\ref{flatband}), with lower band having flatness ratio $3.13$. (middle) Berry curvature of its lower band, which integrates to a Chern number of $-1$. (right) The matrix elements $d_{00}=\langle 1,0|d_z|0,0\rangle$, $d_{01}=\langle 1,\pm 1|d_\pm|0,0\rangle$, $\mu_{01}=\langle 1,\pm 1|d_\pm |1,0\rangle$, $d_{1}=\langle 1,\pm 1|d_z|1,\pm 1\rangle$, $\mu_0=\langle 1,0|d_z|1,0\rangle$, $d_{0}=\langle 0,0|d_z|0,0\rangle$ (Dark Red, Red, Pink, Orange, Green, Brown) as a function of $\eta$, in units of $d$. The diagonalization was performed over the $\tilde J=0,1$ and $2$ manifold. }
\label{Fig:coldatom}
\end{figure}

From Eq.~(\ref{Hddd}), the electric field modifies the single-body eigenbasis of each molecule via
\begin{equation}
H_d=\frac{\hbar^2}{2I}\hat J^2-Ed\cos\theta=\frac{\hbar^2}{2I}\left(\hat J^2-\eta\cos\theta\right)
\end{equation}
where $\eta=\frac{2IEd}{\hbar^2}$ is the dimensionless tunable ratio between the rotational energy scale and the dipole energy. In the basis of "bare" angular momentum eigenstates $|\tilde J,\tilde M\rangle$ corresponding to $\vec E=0$, 
\begin{eqnarray}
\langle \tilde J',\tilde M'|H_d|\tilde J,\tilde M\rangle &=&\frac{\hbar^2}{2I}\left(\tilde J(\tilde J+1)\delta_{\tilde J\tilde J'}\delta_{\tilde M\tilde M'}-\sqrt{\frac{4\pi}{3}}\eta\int Y^*_{\tilde J'\tilde M'}Y_{10}Y_{\tilde J\tilde M}d\Omega\right) \notag\\
&=&\frac{\hbar^2}{2I}\left[\tilde J(\tilde J+1)\delta_{\tilde J\tilde J'}\delta_{\tilde M\tilde M'}-\eta\sqrt{(2\tilde J'+1)(2\tilde J+1)}\left(\begin{matrix}\tilde J' & 1 &\tilde J\\ 0 & 0 &0 \end{matrix}\right) \left(\begin{matrix}\tilde J' & 1 &\tilde J\\ \tilde M' & 0 &\tilde M \end{matrix}\right)\right]
\end{eqnarray}
where $\left(\begin{matrix}\tilde J' & 1 &\tilde J\\ \tilde M' & 0 &\tilde M \end{matrix}\right)$ is a 3-j symbol. 
Expressed as these matrix elements, $H_d$ can then be numerically diagonalized to yield the lowest four eigenstates $|0,0\rangle,|1,0\rangle$ and $|1,\pm 1\rangle$ that are adiabatically connected to $|\tilde 0,\tilde 0\rangle,|\tilde 1,\tilde 0\rangle$ and $|\tilde 1,\pm \tilde 1\rangle$. With them, one can next compute the dipole moments $d_{00}$, $\mu_{01}$, etc. by expressing $d_z$ and $d_\pm$ as real-space spherical harmonics $\sqrt{\frac{4\pi}{3}}Y_{10}$ and $\sqrt{\frac{4\pi}{3}}Y_{1,\pm1}$. These results are presented in Fig.~\ref{Fig:coldatom}c as a function of $\eta$.

As presented in Fig.~3 of the main text, very small $\eta$ gives very large inter-LL separation, but will probably not host stable ground states as the dynamical PPs are larger than the static ones, and the interaction as a whole oscillates between being attractive and repulsive. Intermediate values of $\eta\approx 3$ are probably optimal for hosting stable FQH states that rely on 3-body interactions, with dominant static inter-sublattice PPs (of symmetry type $\lambda=[2,1]$) slightly larger than their dynamic counterparts.

\subsection{Construction of flatband model}

We consider a checkerboard lattice, superficially similar to that in Ref~\onlinecite{yao2013realizing}, with nearest-neighbor (NN) hoppings in diagonal directions and next-nearest-neighbor (NNN) hoppings in horizontal and vertical directions, such that $R_{NNN}=\sqrt{2}R_{NN}$. NN hoppings connect dissimilar sublattices A and B, while NNN hoppings connect two A or two B sites. 

Eq.~(\ref{tAB}) gives a 2-component Hamiltonian that can be optimized to produce Chern bands that are approximately flat. After optimization, we obtain an illustrative flat band model with lower band possessing a flatness ratio (bandgap/bandwidth) of $3.13$ and Chern number $-1$:
\begin{eqnarray}
H_{flatband}&=&-(0.9378\cos k_x+0.583\cos k_y)\mathbb{I}-\left(1.2626\cos\frac{k_x}{2}\cos\frac{k_y}{2}+7.6819\sin\frac{k_x}{2}\sin\frac{k_y}{2}\right)\sigma_x\notag\\
&& -\left(4.5566\cos\frac{k_x}{2}\cos\frac{k_y}{2}-2.212\sin\frac{k_x}{2}\sin\frac{k_y}{2}\right)\sigma_y+(1.4037\cos k_x-3.7331\cos k_y)\sigma_z
\label{flatband}
\end{eqnarray}
This local optimum, which is realized with parameters $(\lambda,\phi_0,s_A,s_B,v_A,v_B,\Lambda)=(2.26,0.18,0.09+0.751i,0.612+0.31i,0.08+0.928i,-0.049+0.011i,0.97)$, is different from that in Ref.~\onlinecite{yao2013realizing}, which is not subject to additional constraints from dynamical modulations that do not modulate the hoppings (Eq.~\ref{WAB}). As such, we obtained flatness ratio that is slightly lower than that in Ref.~\onlinecite{yao2013realizing}, although probably still sufficiently flat for hosting FCI ground states~\cite{kourtis2014fractional}. If desired, the flatness ratio can be further optimized by including further hoppings, or by considering optical lattice tensor shifts on the different sublattices.

\newpage

\end{document}